\documentclass[manuscript]{aastex}

\slugcomment{Submitted to Icarus}

\shorttitle{HST/WFC3 Observations of Uranus' 2014 Storm Clouds}
\shortauthors{Irwin et al.}

\begin{document}


\title{HST/WFC3 Observations of Uranus' 2014 storm clouds and comparison with VLT/SINFONI and IRTF/SpeX observations}


\author{Patrick G. J. Irwin}
\affil{Department of Physics, University of Oxford, Parks Rd, Oxford OX1 3PU, UK.}
\email{patrick.irwin@physics.ox.ac.uk}

\author{Michael H. Wong}
\affil{University of California at Berkeley Astronomy Department, Berkeley, CA 947200-3411, USA}

\author{Amy A. Simon}
\affil{NASA Goddard Space Flight Center Solar System Exploration Division (690), Greenbelt, MD 20771, USA}

\author{G.S. Orton}
\affil{Jet Propulsion Laboratory, California Institute of Technology, 4800 Oak Grove Drive, Pasadena, CA 91109, USA.}

\and

\author{Daniel Toledo}
\affil{Department of Physics, University of Oxford, Parks Rd, Oxford OX1 3PU, UK.}



\begin{abstract}
In November 2014 Uranus was observed with the Wide Field Camera 3 (WFC3) instrument of the Hubble Space Telescope as part of the Hubble 2020: Outer Planet Atmospheres Legacy program, OPAL. OPAL annually maps Jupiter, Uranus and Neptune (and will also map Saturn from 2018) in several visible/near-infrared wavelength filters. The Uranus 2014 OPAL observations were made on the 8/9th November at a time when a huge cloud complex, first observed by \cite{dePater15} and subsequently tracked by professional and amateur astronomers \citep{sayanagi16}, was present at 30 -- 40$^\circ$N. We imaged the entire visible atmosphere, including the storm system, in seven filters spanning 467 -- 924 nm, capturing variations in the coloration of Uranus' clouds and also vertical distribution due to wavelength dependent changes in Rayleigh scattering and methane absorption optical depth. Here we analyse these new HST observations with the NEMESIS radiative-transfer and retrieval code in multiple-scattering mode to determine the vertical cloud structure in and around the convective storm cloud system.

The same storm system was also observed in the H-band (1.4 -- 1.9 $\mu$m) with the SINFONI Integral Field Unit Spectrometer on the Very Large Telescope (VLT) on 31st October and 11th November, reported by Irwin et al. (2016, 10.1016/j.icarus.2015.09.010). To constrain better the cloud particle sizes and scattering properties over a wide wavelength range we also conducted a limb-darkening analysis of the background cloud structure in the 30 -- 40$^\circ$N latitude band by simultaneously fitting: a) these HST/OPAL observations at a range of zenith angles; b) the VLT/SINFONI observations at a range of zenith angles; and c) IRTF/SpeX observations of this latitude band made in 2009 at a single zenith angle of 23$^\circ$, spanning the wavelength range 0.8 -- 1.8 $\mu$m (Irwin et al., 2015, 10.1016/j.icarus.2014.12.020).

We find that the HST observations, and the combined HST/VLT/IRTF observations are well modelled with a three-component cloud comprised of: 1) a vertically thin, but optically thick `deep' tropospheric cloud at a pressure of $\sim 2$ bars; 2) a methane-ice cloud at the methane-condensation level with variable vertical extent; and 3) a vertically extended tropospheric haze. We find that the particles sizes in both haze and tropospheric cloud have an effective radius of $\sim 0.1$ $\mu$m, although we cannot rule out larger particle sizes in the tropospheric cloud. 
We find that the particles in both the tropospheric cloud and haze are more scattering at short wavelengths, giving them a blue colour, but are more absorbing at longer wavelengths, especially for the tropospheric haze. For the particles in the storm clouds, which we assume to be composed of methane ice particles, we find that their mean radii must be $\sim 0.5$ $\mu$m. We find that the high clouds have low integrated opacity, and that ``streamers'' remininiscent of convective thunderstorm anvils are confined to levels deeper than 1 bar. These results argue against vigorous moist convective origins for the cloud features.
\end{abstract}





\section{Introduction}

Long-term observations of the outer planets are critical to understanding the atmospheric dynamics and evolution of gas giant planets \citep{visions11}.
To this end, the Hubble 2020: Outer Planet Atmospheres Legacy (OPAL) program provides for yearly outer planet monitoring using the Wide Field Camera 3 (WFC3) for the remainder of Hubble's lifetime. The program began in Hubble Observing Cycle 22, observing Uranus in November 2014 \citep{wong15}, Jupiter in early 2015 \citep{simon15} and Neptune in September 2015 \citep{simon16, wong16}. 

The November 2014 Uranus observations, reported here, occurred at a time when a large cloud feature was observable at 30 -- 40$^\circ$N in Uranus' atmosphere.  Large, bright clouds had previously been observed with the Keck Telescope in August 2014 \citep{dePater15}, and a bright storm cloud system was subsequently detected by amateur observers in September 2014. However, the bright cloud seen by amateurs was not the very bright `Br' feature seen by \cite{dePater15}, but instead seemed to have evolved from a fainter `Feature 2', seen at 33$^\circ$N. These detections prompted a number of ground-based Director's Discretionary Time (DDT) observations at the world's leading observatories, including observations in the H-band (1.4--1.8 $\mu$m) made with the SINFONI integral field unit spectrograph at the European Southern Observatory (ESO) Very Large Telescope (VLT) by \cite{irwin16}. The evolution of the storm features was tracked by both professional and amateur imaging observations over the 2014--2015 time period \citep{sayanagi16}.

The VLT/SINFONI observations were made with adaptive optics and have a spatial resolution of $\sim$0.1\arcsec. However, since Uranus' disc has a diameter of only $\sim 3.6$\arcsec  this translates to only $\sim 36$ resolution elements from limb to limb. Analysing the spectral observations, \cite{irwin16} found that the background spectra at 30 -- 40$^\circ$N could be modelled with an optically thick, but vertically thin cloud based at the $\sim 2$-bar level combined with one or two haze layers in the upper troposphere/lower stratosphere and a methane cloud based at 1.23 bar and with variable vertical extent. With the spatial resolution available, the main cloud appeared to be sheared with height with the deeper part of the cloud to the southwest and the higher part to the northeast.

In this paper we compare the new HST/WFC3 observations made in just seven filter channels spanning 467 -- 924 nm, but with extremely high spatial resoltion (0.05\arcsec resolution at 600 nm, with 0.04\arcsec pixel size in the WFC3/UVIS detector) made within just a few days of the lower spatial resolution, but high spectral resolution VLT/SINFONI observations spanning the longer wavelength 1.436 -- 1.863 $\mu$m spectral range. Using these data we attempt to find aerosol vertical distributions that are simultaneously consistent with both these observations and also earlier IRTF/SpeX Uranus observations, made in 2009 \citep{tice13}.

\section{Observations}

\subsection{HST/WFC3 observations}
Observations of Uranus were made with the Wide Field Camera 3 (WFC3) of the Hubble Space Telescope (HST) in seven spectral channels, listed in Table \ref{tbl-1}, during eight HST orbits on November 8--9th 2014, spanning 32.6 hours and 680 deg of central meridian longitude (1.9 Uranus rotations). The data were navigated by aligning a simulated (limb-darkened and PSF-convolved) Uranus disc to the image \citep{Lii10}. The listed photometric errors are combined from several factors: uncertainty in the solar spectrum \citep{colina96}, accuracy of the photometric calibration of the WFC3 instrument \citep{dressel16}, and uncertainty in the correction for fringing in narrowband filter images at wavelengths longer than $\sim 675$ nm \citep{wong11}. We also estimate that the failure to include quad filters in the major photometric calibration update for WFC3 in 2016 \citep{ryan16} adds a 3.1\% uncertainty.

\begin{figure}
\epsscale{1.0}
\plotone{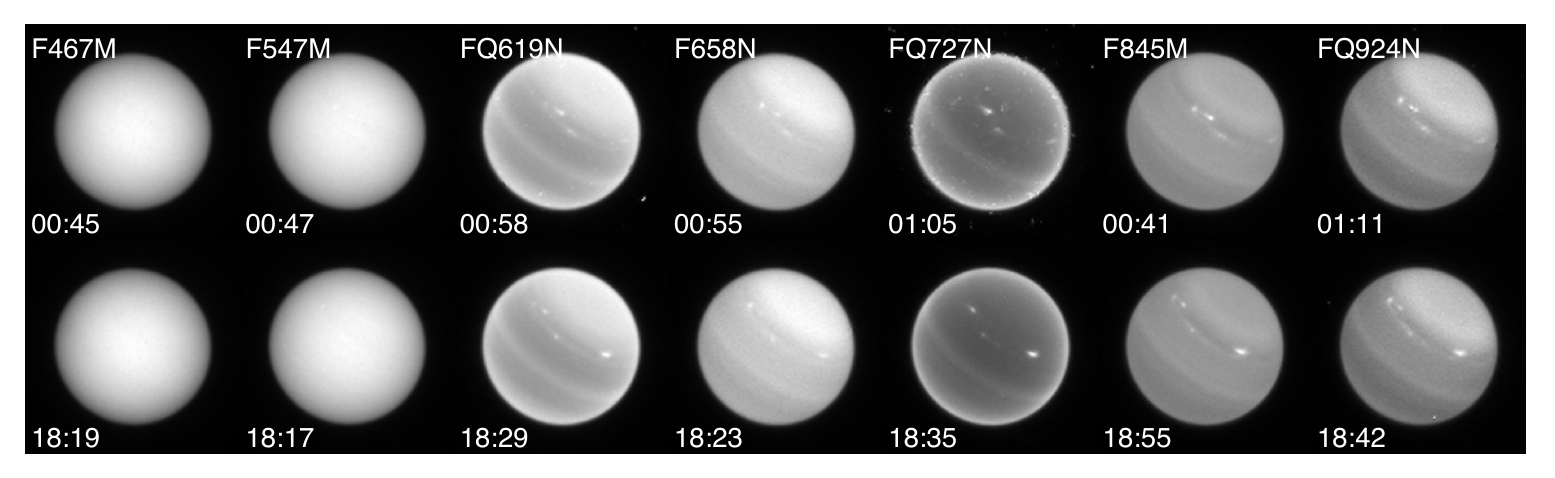}
\caption{HST/WFC3 observations of Uranus on 9th November 2014 near 01:00 UT (top row) and near 18:30 UT (bottom row).  The filter identity is indicated at the top left of each column, while the time of observations in indicated at the bottom left of each image.\label{uranusnav}}
\end{figure}

\begin{figure}
\epsscale{.80}
\plotone{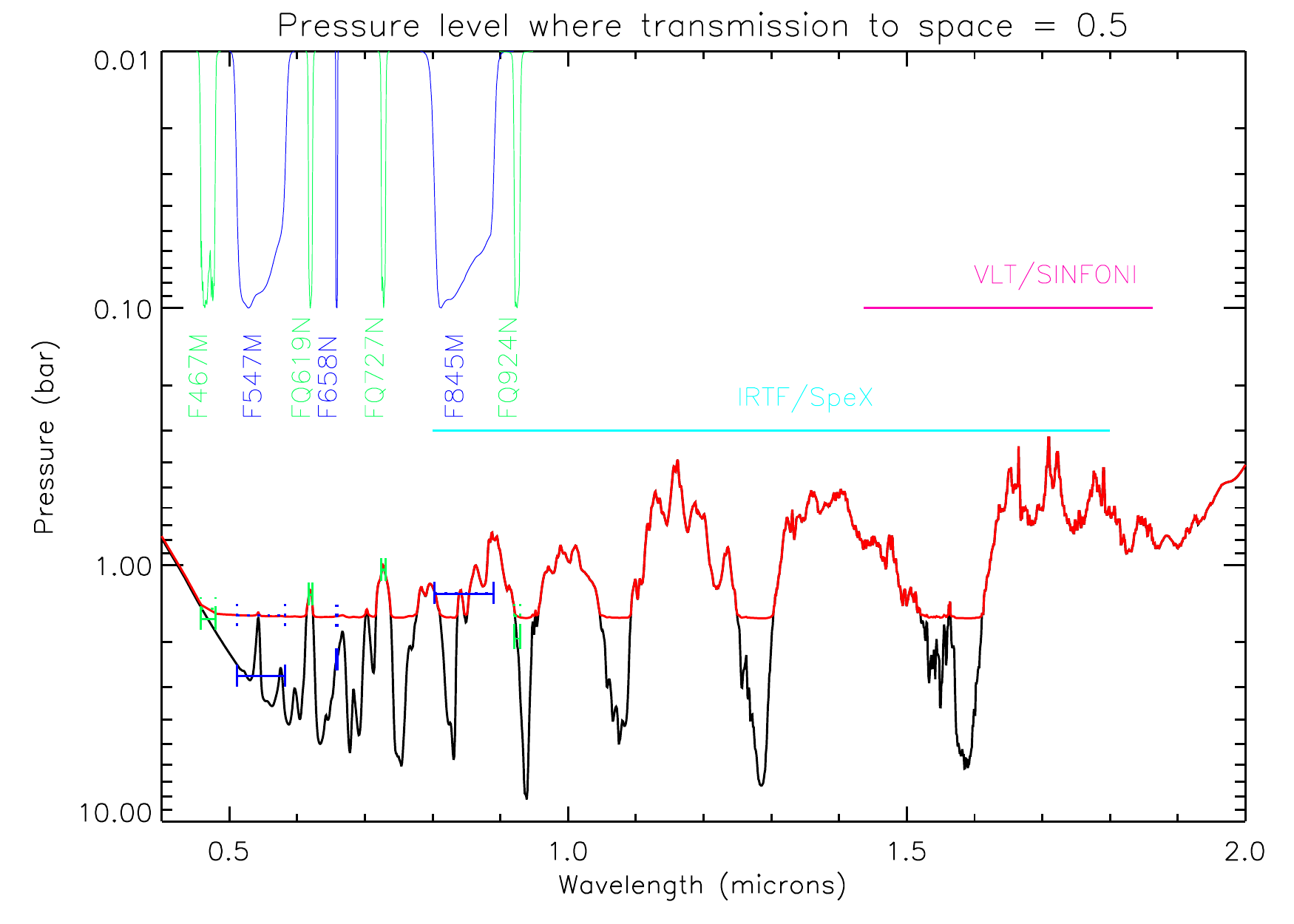}
\caption{Pressure level at which transmission to space is 0.5 for vertical path in our reference Uranus atmosphere at IRTF /SpeX resolution (black line), together with HST/WFC3 Filter profiles and pressure levels where the filter-averaged transmission is 0.5.  The red line is the pressure level where transmission is 0.5 for a typical cloudy atmosphere retrieved from the observations. The pressure levels where the filter-averaged transmission is 0.5 for the cloudy case are marked with dotted lines. Also plotted are the wavelength ranges spanned by the IRTF/SpeX and VLT/SINFONI observations.\label{uranustrans}}
\end{figure}

Figure \ref{uranusnav} shows the appearance of Uranus in the seven channels at around 01:00 UT on 9th November 2014, with the storm cloud system near the centre of Uranus' disc, and then later that day (at around 18:30 UT), with the main storm clouds nearer the evening terminator (i.e. upper left) and a trailing cloud more visible near the morning terminator. The levels probed in these filters is shown in Fig. \ref{uranustrans}, where we show the pressure level in our reference Uranus atmosphere (described in section 3) where the vertical transmission to space (calculated assuming the IRTF/SpeX spectral resolution of 0.002 $\mu$m) is 0.5, together with the HST/WFC3 filter profiles (and filter-averaged pressure levels where the transmission to space is 0.5). In the cloud-free case the atmospheric extinction is dominated by gaseous Rayleigh scattering at short wavelengths and gaseous methane absorption bands that become increasingly stronger at longer wavelengths. There is also an H$_2$--H$_2$ collision-induced-absorption (CIA) band at $\sim$820 nm.  

Figure \ref{uranustrans} also shows the 0.5-transmission level for an atmosphere with clouds necessary to match the observed spectra typical of the storm cloud latitude system (Section 3.4). Although many transparent gas-absorption windows within our HST/IRTF/VLT bandpass are shown by dips in the clear atmosphere case (black line), the deepest levels probed are actually limited to $p < 2$ bar by the presence of cloud layers (red line). Figure \ref{uranusnav}  reveals a remarkable difference between the appearance of Uranus at wavelengths shorter and longer than $~600$ nm. Images in Fig. \ref{uranusnav} at 467 nm (blue) are largely featureless, while images at 658 nm (red) and longer show significant contrast between cloud features and the background atmosphere. However, Fig. \ref{uranustrans} suggests that the red and blue filters both reach (two-way) unit optical depth at about the same level of 1.5 bar. The difference in appearance can be attributed to the dominance of scattering in the opacity in the blue channel, and the dominance of absorption in the red channel. Thus, cloudy regions present weak contrast in the scattering-dominated blue channel, where Rayleigh scattering is so strong that additional scattering from clouds is not a large effect.

From Fig. \ref{uranusnav} we can see that the appearance of Uranus in the Rayleigh-scattering-dominated channels at 467 and 547 nm is somewhat featureless, but features start to emerge as we go to longer wavelengths and the Rayleigh scattering opacity falls off as $1/\lambda^4$. In addition to the storm clouds themselves, which are more fully described below, we can see that a northern polar `hood' of enhanced reflection is seen at the longer wavelengths of low methane transmission poleward of $\sim 45^\circ$N (planetographic latitudes are assumed throughout), which must be at a similar pressure level to the main clouds. We can also see a region of enhanced reflectivity over the equator, which as it can be seen at wavelengths of strong methane absorption must extend to, or lie well above the main cloud deck.

\begin{figure}
\epsscale{0.7}
\plotone{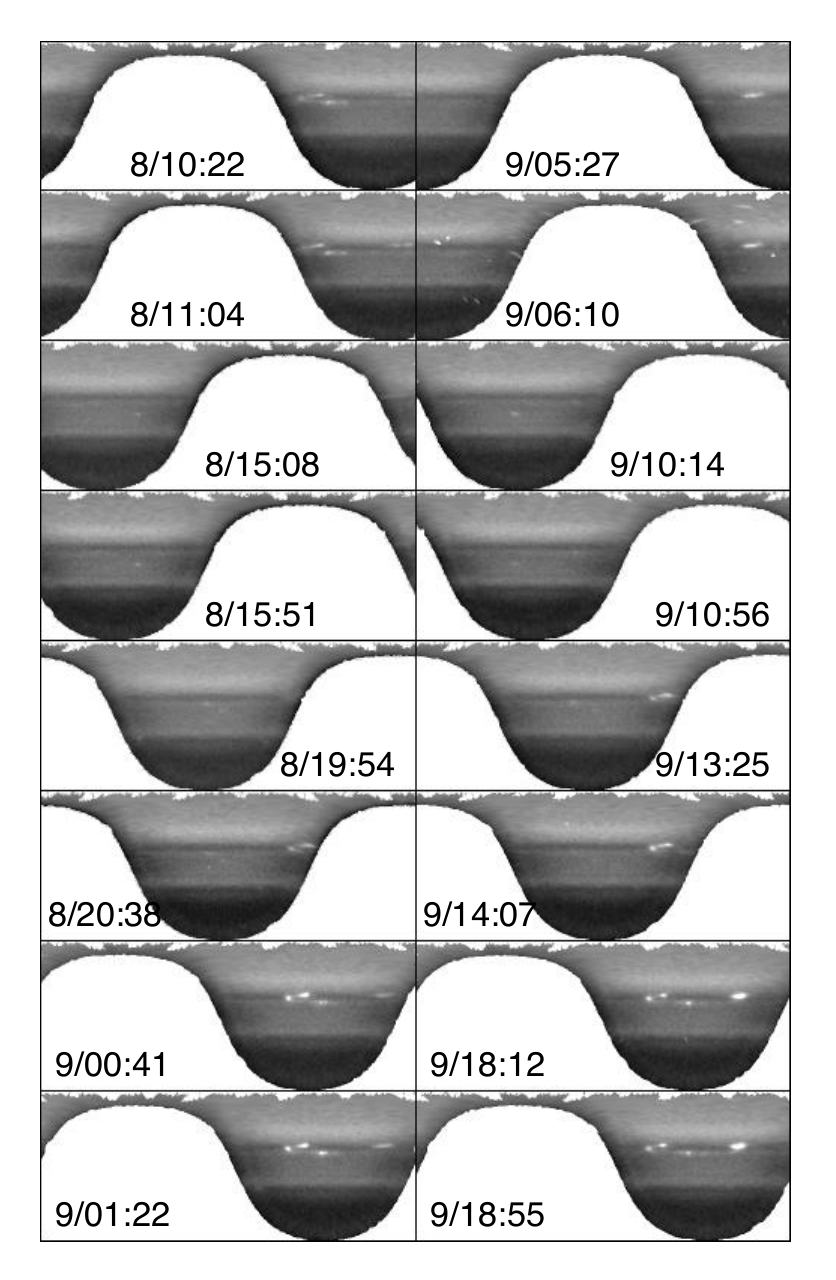}
\caption{All HST/WFC3 observations made with F845M filter on 8/9th November 2014 projected onto a rectangular latitude/longitude grid spanning the range 90$^\circ$S to 90$^\circ$N, and 0 -- 360$^\circ$E. The observations show the storm system moving slowly from right to left as time progresses.   \label{uranus845summary}}
\end{figure}

\begin{figure}
\epsscale{0.45}
\plotone{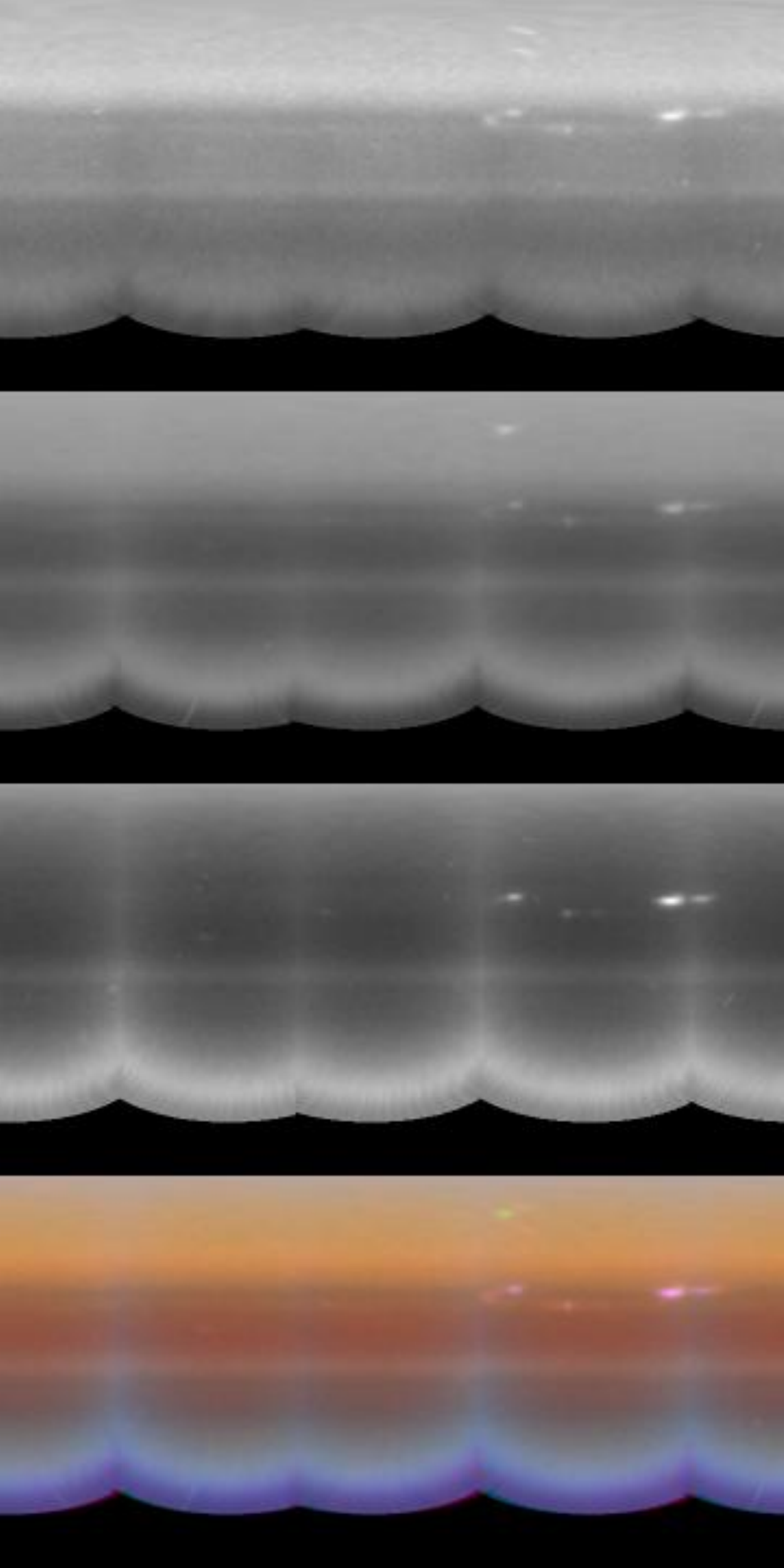}
\caption{Average of all observations made 9th November 2014 in: 1) F924N (weak methane absorption - top row); 2) FQ619N (medium methane absorption - second row); and 3) FQ727N (strong methane absorption - third row) filters, co-projected on to a rectangular latitude/longitude grid (90$^\circ$S -- 90$^\circ$N, 0 -- 360$^\circ$E). The F924N filter shows reflection from clouds at all observable levels. From Fig.\ref{uranustrans} we can see that the FQ619N shows reflection from clouds based at pressures less than $\sim$ 1.2 bars, while the  FQ727N filter shows reflection from clouds lying at pressures less than $\sim$ 1 bar. The bottom panel shows a colour composite, with F924N (red), FQ619N (green) and FQ727N (blue). In this scheme, deep clouds are red, medium altitude clouds are yellow and very high clouds are blue-green to white.  \label{uranus_false_color}}
\end{figure}

All observations of Uranus made with the F845M filter, which probes to the deepest observable clouds and have high signal-to-noise, are shown in Fig. \ref{uranus845summary}, projected on to a rectangular latitude/longitude frame. 
Figures \ref{uranusnav} and \ref{uranus845summary} show that the main storm system actually consists of three clouds: 1) a `deep' cloud visible only in filters probing to pressures $>$ $\sim1$ bar; 2) a bright, high cloud to the northeast visible at all wavelengths longer than 547 nm; and 3) a slightly less bright, high cloud to the southeast, which in Fig. \ref{uranusnav} seems to reside in a generally brighter cloud lane.  The later observations near 18:30UT on 9th November also show a third bright, high cloud at the same latitude as the deep cloud, but trailing $\sim 100^\circ$ to the east. Figure \ref{uranus_false_color} shows all the observations made on 9th November 2014 projected onto the rectangular latitude/longitude grid and combined (N.B. Full rectilinear maps at all wavelengths are available at https://archive.stsci.edu/prepds/opal/). Three filters are shown: F924N, FQ619N and FQ727N, which from Fig.\ref{uranustrans} can be seen to experience increasing methane opacity and are thus only sensitive to clouds at increasingly higher altitudes. The bottom panel of Fig.\ref{uranus_false_color} shows a false-colour composite of the three images with F924N (red), FQ619N (green) and FQ727N (blue); in this representation, deep clouds appear red, and high clouds appear white. Since these observations are averages over a whole day, the differential rotation rates at different latitudes is not completely corrected for, causing some features to be distorted, especially the trailing high cloud $\sim 100^\circ$ to the east (marked as location 6 in Fig.\ref{uranus_false_color}), which here appears to be split into two clouds, but which from Fig.\ref{uranusnav} is actually clearly only one. Hence, it is more instructive to examine the storm system using single observations in these three filters taken within a short time of each other. Figure \ref{uranus_false_color1} shows the appearance of the storm cloud system near 01:00UT on 9th November, while Fig. \ref{uranus_false_color2} shows the appearance later in the day near 18:30UT. From tracking of the storm cloud system \citep{sromovsky14}, the expected System III longitude for the 32$^\circ$N `K1' bright feature on 9th November was 243$^\circ$E near 01:00 reducing to 230$^\circ$E later that day. The main cloud can thus be identified to be the discrete red cloud in the false colour images at 34$^\circ$N, 240$^\circ$E at 01:00UT (Fig. \ref{uranus_false_color1}) and 34$^\circ$N, 220$^\circ$E at 18:30UT (Fig. \ref{uranus_false_color2}). In both observations, the bright cloud to the northeast (which appears blue in this false colour scheme) is at  38$^\circ$N, 255$^\circ$E at 01:00UT and can actually be differentiated as two discrete clouds, with the northeast component being slightly brighter. The bright cloud to the southeast is at 30$^\circ$N, 270$^\circ$E at 01:00UT and there are traces of the trailing high cloud at 38$^\circ$N, 330$^\circ$E. The observations at 18:30UT appear better resolved and show all the same features (although the longitudes have shifted given Uranus' strong retrograde zonal wind speeds of $\sim 100$ m/s at these latitudes \citep{hammel01})  and show that the southeast bright, high cloud appears to reside in a generally brighter cloud `lane' at 30$^\circ$N. There are also traces of what appears almost like the bow wave of a ship curling to the northeast and southeast from the deep cloud, which runs into the cloud lane at 30$^\circ$N, but seems to dissipate to the northeast. This morphology raises the intriguing prospect that the bright high clouds may actually have less to do with convective storm clouds and more to do with `orographic' uplift of air flowing over and around the deep cloud (where we expect the zonal wind speeds to be less) and causing gravitational waves in a region of stably stratified air in the upper troposphere. It is conceivable that the trailing cloud $\sim100^\circ$ to the east may also be dynamically linked in some way. It is difficult to discern any definite temporal changes in the clouds over the two days of observations, due to the varying central meridian longitudes of the different observations and thus different viewing geometries.
   
\subsection{VLT/SINFONI observations}
Observations of Uranus were made with the SINFONI instrument on October 31st  and November 11th 2014 at the European Southern Observatory (ESO) Very Large Telescope (VLT) in La Paranal, Chile, previously reported by \cite{irwin16}.

SINFONI is an Integral Field spectrograph that can make use of adaptive optics to yield a spatial resolution of typically 0.1\arcsec and returns $64 \times 64$ pixel `spectral cubes', where each element is a spectrum with 2048 wavelengths. SINFONI has three pixel scale settings: 0.25\arcsec , 0.1\arcsec  and 0.025\arcsec  giving Fields of View (FOV) of  8\arcsec  $\times$ 8\arcsec, 3\arcsec  $\times$ 3\arcsec  and 0.8\arcsec  $\times$ 0.8\arcsec, respectively. Uranus was observed using the 0.1\arcsec pixel scale and the H--grism, which has a spectral resolution of $R=\lambda/\Delta\lambda \sim 3000$ and covers the wavelength range 1.436 -- 1.863 $\mu$m. Since the FOV at the 0.1\arcsec pixel scale was smaller than the apparent disc size of Uranus at this time ($\sim$ 3.7\arcsec) dithered, overlapping observations were recorded on a 2 $\times$ 2 grid, with additional observations centred on Uranus' disc. The data were reduced with the ESO VLT SINFONI pipeline, with additional corrections as described by \cite{irwin16}, and then smoothed to the resolution of IRTF/SpeX ($\Delta\lambda = 0.002 \mu$m) to improve the SNR and also enable easier comparison with the IRTF/SpeX observations.

\begin{figure}
\epsscale{0.6}
\plotone{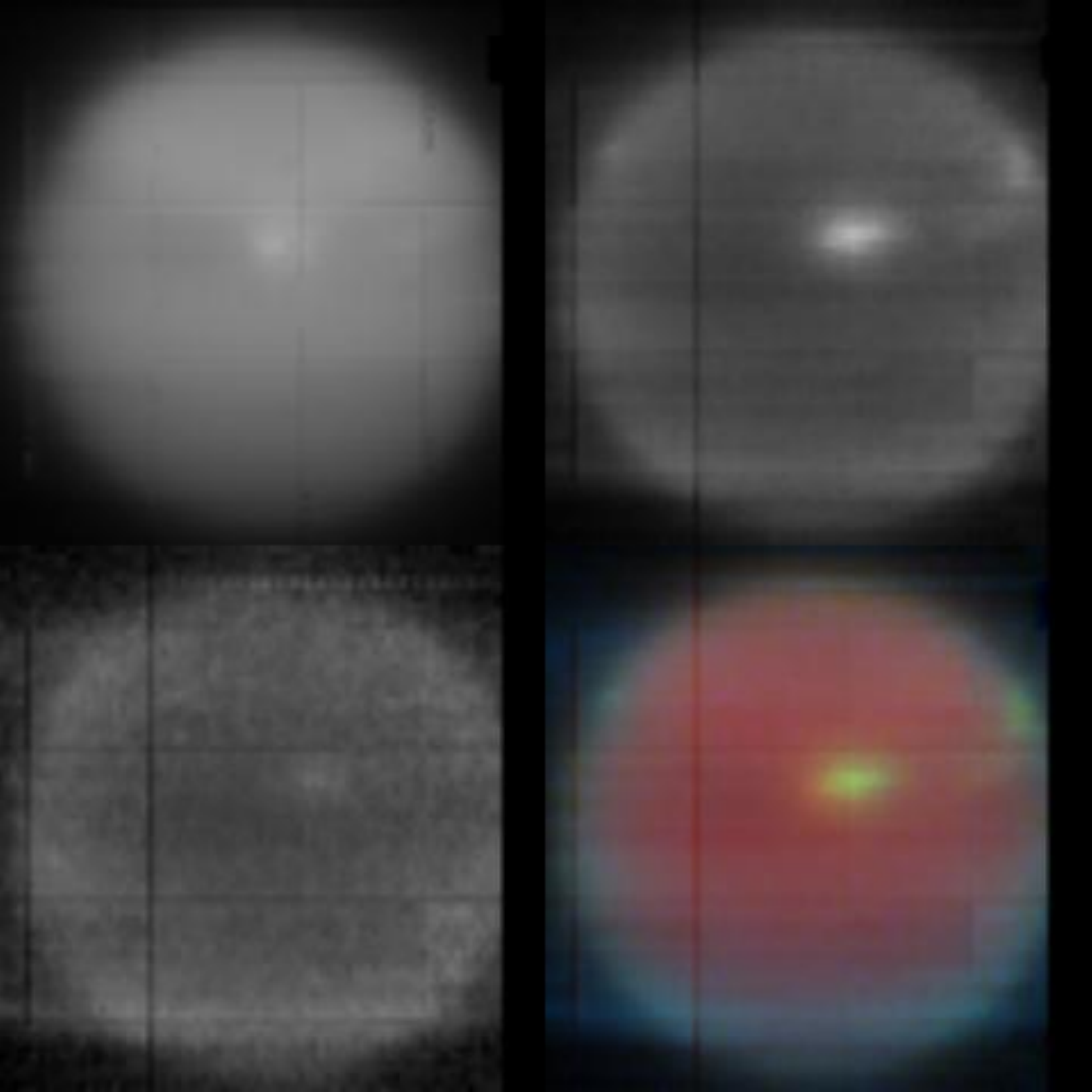}
\caption{Observations on 11th November 2014 with VLT/SINFONI in regions of weak, medium and strong methane absorption, and colour composite (after \cite{irwin16}).  Methane absorption is much stronger at these wavelengths and the pressure levels sounded in the three individual images are $p < 3$ bar (top left), $p <  1.25$ bar (top right) and $p < 0.35$ bar (bottom left) respectively. At these wavelengths it would appear that we are sensitive to the deep cloud at  $34^\circ$N, which again appears red, and its partner to the northeast, which here is shown to extend almost to the tropopause. The trailing partner almost 100$^\circ$ to the east is just visible in these images, but was not observed at all on 31st October \citep{irwin16}, suggesting that it formed sometime in the intervening 12 days.  \label{uranus_false_color_VLT}}
\end{figure}

Observations recorded on 11th November 2014, previously reported by \cite{irwin16} are shown in Fig. \ref{uranus_false_color_VLT}. Here, three images are shown, which are averages of wavelengths with weak (red), medium (green) and high (blue) methane absorption. The false colour composite has the same characteristics as the HST observations shown in Figs \ref{uranus_false_color} -- \ref{uranus_false_color2}, with deep clouds appearing red and high clouds white. Comparing Fig. \ref{uranus_false_color_VLT} and Figs \ref{uranus_false_color} --  \ref{uranus_false_color2} in the area of the storm system, we can see that the apparent shearing of cloud centre with altitude reported by \cite{irwin16}  is in the HST observations spatially resolved to in fact be two clouds with fixed position relative to each other, but at different altitudes. It is also clear that both the HST and VLT observations detect the trailing bright, high cloud $\sim 100^\circ$ to the east.

\subsection{IRTF/SpeX observations}
Long-slit spectral observations Uranus were made in 2009 with the SpeX instrument on NASA's Infrared Telescope Facility (IRTF) on Mauna Kea, Hawaii. As reported by \cite{tice13} the slit was aligned with
Uranus' central meridian and spectra recorded from 0.8 to 1.8 $\mu$m with a spectral resolution of $R=\lambda/\Delta\lambda = 1200$ and with an average `seeing'  that varied from 0.5\arcsec in the H-band (1.4 -- 1.8 $\mu$m) to 0.6\arcsec in the I-band (0.8 -- 0.9 $\mu$m). The observations were reanalysed by  \cite{irwin15}, who found that the observed spectrum was consistent with one of either two models: A) a simple two-cloud model, consisting of a vertically-thin `deep' cloud near the 2-bar level, together with a vertically extended tropical haze; or B) a modified form of the more complicated 5-component model of \cite{sromovsky11}, consisting three vertically thin clouds in the lower troposphere, together with a vertically extended tropospheric haze from 1 -- 0.1 bar, and a vertically extended stratospheric haze from 0.1 -- 0.01 bar. The three lower clouds of the \cite{sromovsky11} model were based at $\sim 5$bar, $\sim 2 - 3$bar and $\sim 1$ bar, the upper of which was interpreted to be composed of methane ice and set to the condensation level of 1.2 bar. \cite{irwin15} found that both models could be made to fit the IRTF/SpeX spectra well, although they had a slight preference for the latter. Although recorded five years before the VLT/SINFONI and HST/WFC3 observations, these IRTF/SpeX data from 2009 are remarkably consistent with the latter observations in regions away from discrete clouds and so we use them here, together with the later observations, to gain a better understanding of the background atmospheric state.

\section{Analysis of background cloud atmospheric state}

The HST/WFC3 observations are comprised of just seven filter-averaged observations and thus individual sets of observations contain at most seven pieces of independent information on the vertical distribution of clouds in Uranus' atmosphere.
Since to describe fully the vertical distribution of cloud opacity and also determine the cloud particles sizes and scattering properties requires many more than seven pieces of information it is clear that we must first constrain the analysis of the HST/WFC3 observations by constructing a parameterised model of Uranus' clouds that can be represented with only a few variables. We could do this for the HST/WFC3 observations alone, but limiting the analysis to a restricted wavelength range (and a restricted set of discrete wavelengths also) could easily lead to solutions that might be consistent with the HST/WFC3 data, but which are inconsistent with observations at other wavelengths. Since for this analysis we have observations from HST, VLT and IRTF covering the spectral range 0.467 -- 1.8 $\mu$m at multiple zenith angles we have a unique opportunity to constrain the vertical distribution and properties of Uranus' clouds and hazes more reliably than has ever been attempted before. We thus combined all our observations of the background atmosphere at  30 -- 40$^\circ$N, away from the discrete cloud features, and sampled them at a range of zenith angles from near-nadir to near the limb and attempted to construct a cloud model that would simultaneously be consistent with all these observations. Once this background atmospheric state had been fitted we then went on to explore what needed to be added or subtracted to account for the HST/WFC3 and VLT/SINFONI observations of the discrete clouds. To make this quantitative analysis we used the NEMESIS \citep{irwin08} radiative-transfer and retrieval code. We set up our retrieval model as described below.

\subsection{Temperature/Abundance Profiles}

The temperature and abundance profile assumed in this study was the same as that used by \cite{irwin16} and \cite{irwin15}. The temperature profile was based on the `F1' profile determined by \cite{sromovsky11} which has an He:H$_2$ ratio of 0.131 and assumes a 0.04\% mole fraction of neon and a deep CH$_4$ mole fraction of 4\% (reducing with height), after \cite{kark10}.
 
 \subsection{Gaseous Absorption data and Scattering Radiative Transfer Model}
 
Since in this study we analyse observations recorded between 0.467 and 2.0 $\mu$m, it was not appropriate to use the methane lines of the WKMC-80K line database \citep{campargue12}, which only cover the 1.26 -- 1.71 $\mu$m range.  Instead, we used the methane absorption coefficients of \cite{kark10}, which is a consistent set  of absorption coefficients covering the entire spectral region under investigation, although \cite{irwin12} have shown that these coefficients do not model the 1.4 -- 1.8 $\mu$m range as accurately as the WKMC-80K line database. The spectra were fitted with NEMESIS \citep{irwin08}, using a correlated-k model \citep{lacisoinas91} and k-tables generated from the  \cite{kark10} methane absorption coefficients  assuming the IRTF/SpeX triangular instrument function with FWHM =  0.002 $\mu$m and step of 0.001 $\mu$m. In addition, HST-filter-averaged k-tables were computed to model the HST filter observations. For H$_2$--H$_2$ and H$_2$--He collision-induced absorption (CIA) we used the coefficients of \cite{borysow89,borysow00} and \cite{zhengborysow95} and an equilibrium ortho/para-H$_2$ ratio was assumed at all altitudes and latitudes. In addition to H$_2$--H$_2$ and H$_2$--He CIA, H$_2$--CH$_4$ and CH$_4$--CH$_4$ collision-induced absorption was also included \citep{borysowfrommhold86, borysowfrommhold87}. The spectra were simulated using a Matrix Operator multiple scattering code, based on the method of \cite{plass73}, including the Rayleigh scattering by the air molecules themselves, with 5 zenith angles (with Gauss-Lobatto calculated ordinates and weights) and $N$ Fourier components to cover the azimuth variation, where $N$ is set adaptively from the viewing zenith angle, $\theta$, as $N$ = int$(\theta/3)$. A nine zenith angle Gauss-Lobatto quadrature scheme was also tried, but was found to give negligibly improved results and was much slower. To perform this calculation the reference temperature, pressure and abundance profiles were split into 39 levels equally spaced in log pressure between 11 bar and 0.01 bar. 

\subsection{Cloud Models}

The HST observations are in seven channels. As such, we concluded that the five-cloud model of \cite{sromovsky11} had too many free parameters and hence elected to use a more simple, three-cloud model comprising: 1) a `deep' thick tropospheric cloud based near the $\sim 2$ bar pressure level; 2) a methane cloud at the methane condensation level of our reference atmosphere at 1.23 bar; and 3) a tropospheric `haze' cloud, which here we based at the same level as the main cloud deck, but allowed to be vertically extended. We set the \textit{a priori} base of the deep cloud to a level consistent with the level of the main cloud deduced from VLT/SINFONI observations \citep{irwin16} to be 1.9 bar and fixed its \textit{a priori} fractional scale height to various values in the range 0.01 to 0.1, allowing only its opacity to vary. For the methane cloud, we allowed both its fractional scale height and overall opacity to vary. For the tropospheric haze, we set its base to be the same as the deep cloud and fixed its fractional scale height to 1.0, allowing only its opacity to vary.

The scattering properties of all the cloud particles were computed using Mie scattering assuming a standard gamma distribution of particle sizes, although we used Henyey-Geenstein fits to the computed phase functions to average over the characteristic `glory' and `rainbow' of spherical particles, which would not be appropriate for modelling the behaviour of ice particles. For the methane particles we assumed a size distribution with an effective radius of 0.1 or 1.0 $\mu$m and variance 0.1 and for the complex refractive index spectrum we used that found by \cite{martonchik94}. For the `deep' tropospheric cloud particles we tried various \textit{a priori} combinations and eventually chose an effective radius of 0.1 or 1.0 $\mu$m with fixed variance 0.1 and assumed the imaginary refractive index was in the range $(10^{-3} - 10^{-1})$ ($\pm 10\%$) at all wavelengths. In the retrievals we fitted the imaginary refractive index spectrum and then computed the real refractive index spectrum using the Kramers-Kronig relation \citep{sheik05}, assuming the real part of the refractive index to be 1.4 at 467 nm (broadly consistent with the real refractive index of ammonia and methane in the visible).  For the haze we chose an effective variable radius of 0.1 $\mu$m with fixed variance 0.1 and fitted the imaginary refractive index spectrum (and inferred the real part) as just described for the tropospheric cloud particle retrievals.  For the retrieval of the imaginary refractive index spectra of both the tropospheric cloud and tropospheric haze we assumed a correlation wavelength of 0.05 $\mu$m to ensure the retrieved imaginary wavelength spectrum was smooth.

\subsection{Limb-darkening Analysis}

To determine the cloud characteristics over a wide wavelength range and over a range of zenith angles in the latitude band of the storm system, all available observations in the latitude band 30 -- 40$^\circ$N were analysed at a range of zenith angles.

\begin{figure}
\epsscale{0.3}
\plotone{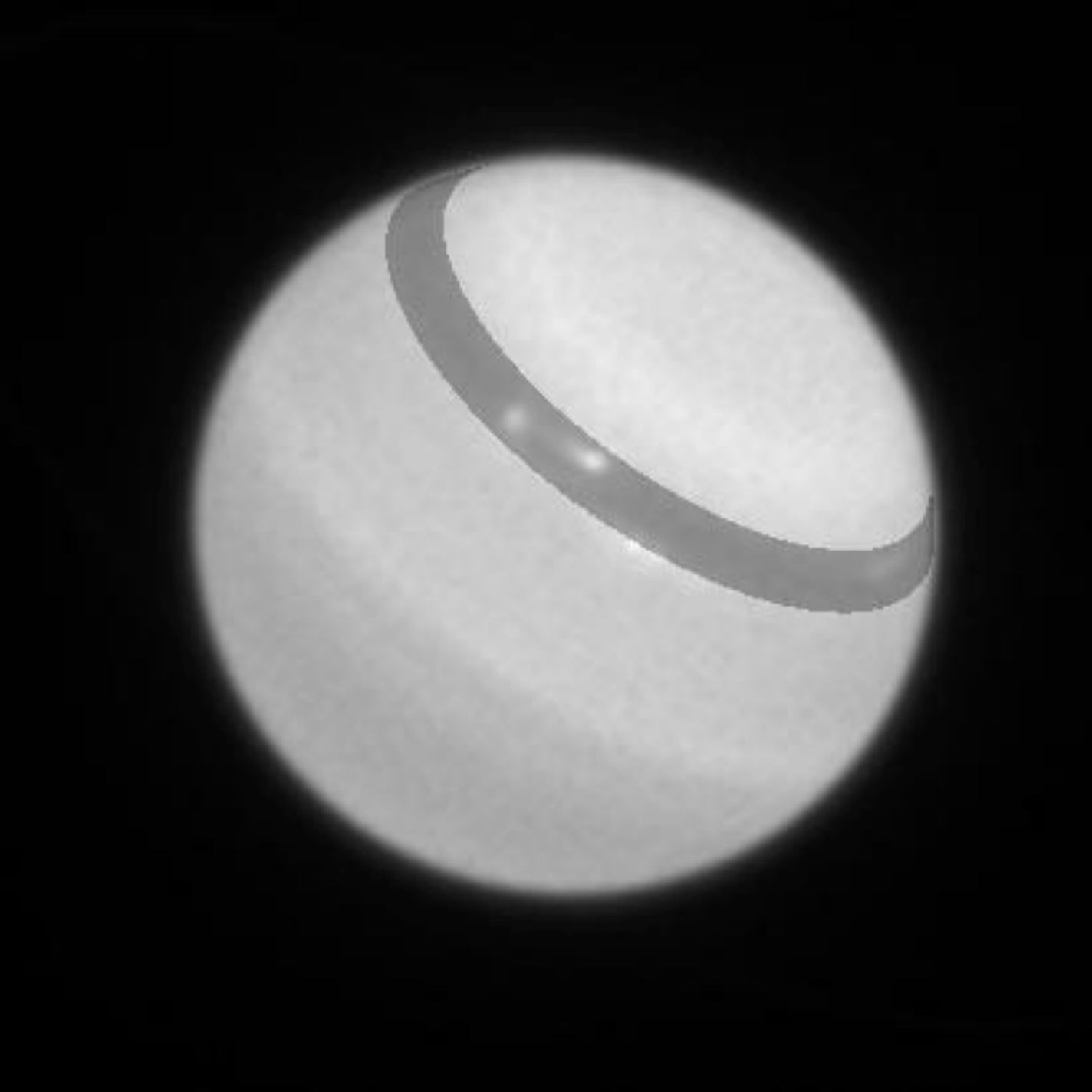}
\caption{HST/WFC3 observation on 9th November 2014 in F845M filter (00:41UT), together with latitude range selected for limb-darkening analysis (30 -- 40$^\circ$N, shaded grey).   \label{uranus_swath}}
\end{figure}

\begin{figure}
\epsscale{0.3}
\plotone{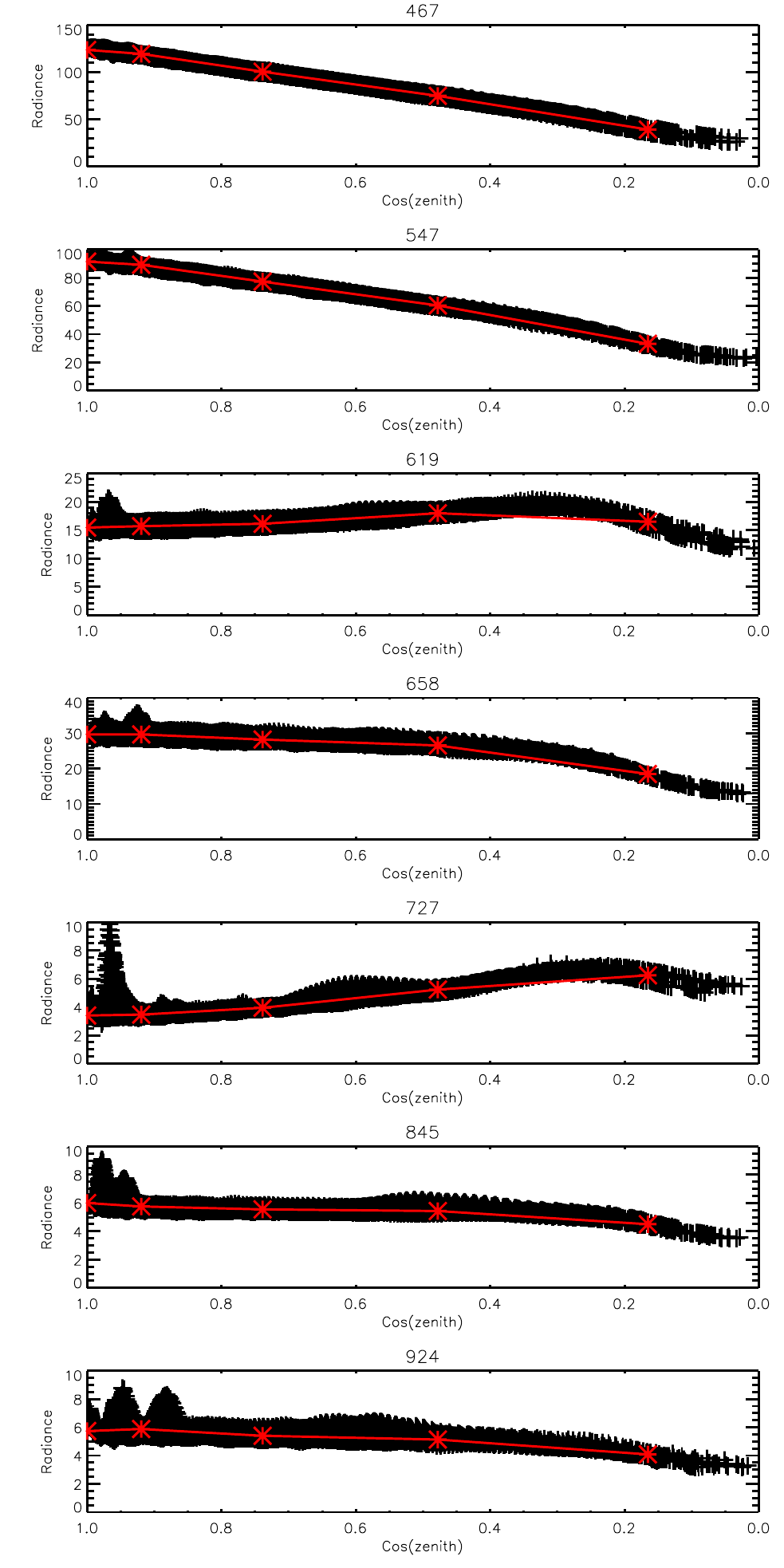}
\caption{Observed HST/WFC3 limb-darkening curves on 9th November 2014 in all seven filters from 00:41 -- 01:11UT for  30 -- 40$^\circ$N, together with cos(zenith) angles sampled (red).    \label{uranus_limbdark_HST}}
\end{figure}

\begin{figure}
\epsscale{1.0}
\plotone{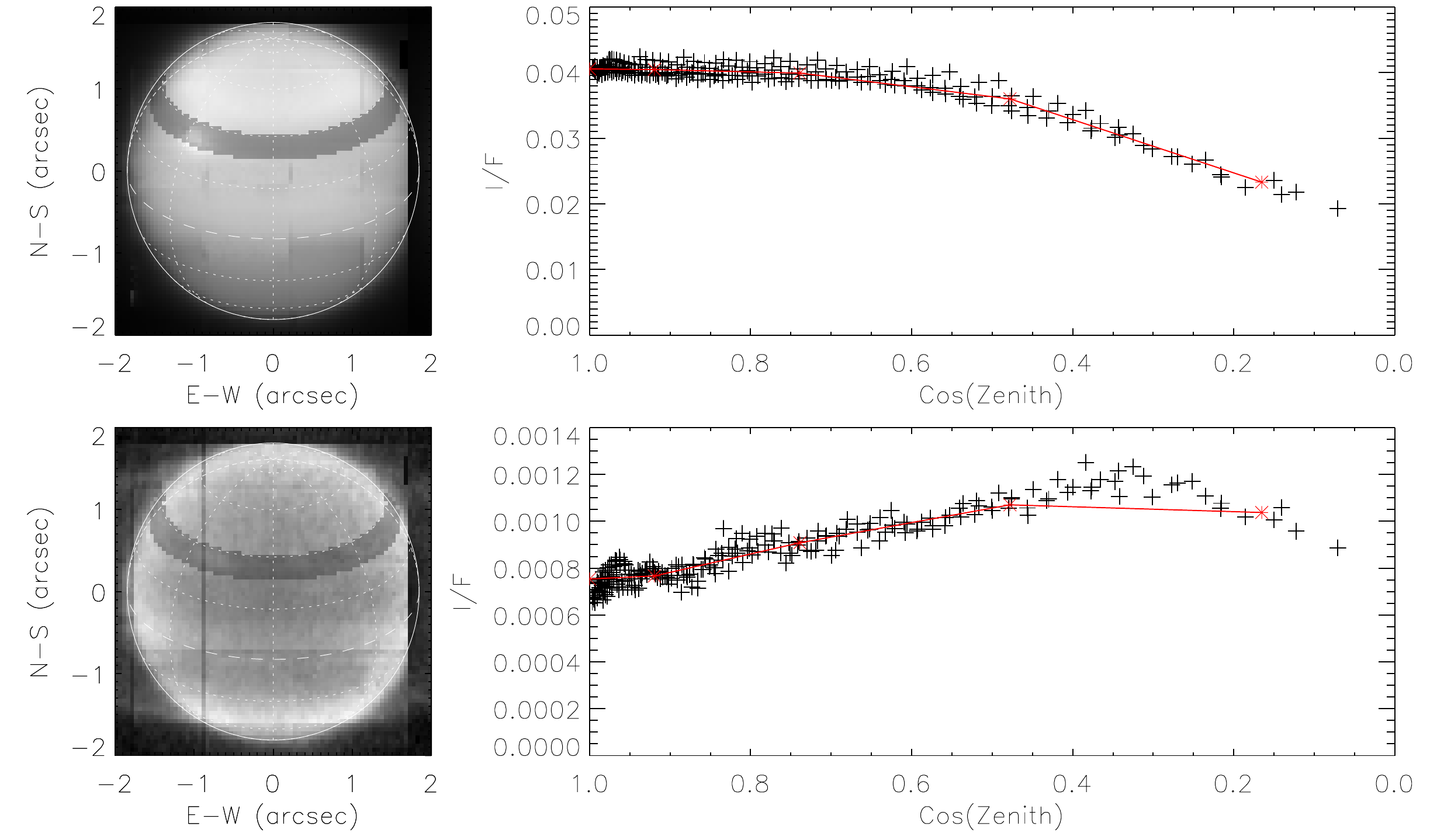}
\caption{Observed VLT/SINFONI limb-darkening curves on 31st October 2014 at 1.59 and 1.683 $\mu$m in latitude range 30 -- 40$^\circ$N, together with cos(zenith) angles sampled (red). Only longitudes to right of central meridian have been sampled to avoid contamination with cloud system locations.  \label{uranus_limbdark_VLT}}
\end{figure}

For HST, the observations near 01:00 UT on  9th November 2014 were used and the latitude swath analysed is shown in Fig.\ref{uranus_swath}. The observations in the seven channels as a function of $\cos(\theta)$, where $\theta$ is the solar zenith angle (assumed to be the same as the emission zenith angle for computational ease), are shown in Fig.\ref{uranus_limbdark_HST}. The observations were averaged and sampled at the five zenith angles used in NEMESIS' Gauss-Lobatto quadrature scheme (listed in Table \ref{tbl-2}) and are overplotted in red. The observations were of sufficient spatial resolution that a median average of all observations was effective in removing storm-affected pixels. A similar analysis was undertaken of the VLT/SINFONI observations. Here we chose one of the observations made on 31st October 2014, which has the best spatial resolution. The appearance of Uranus at two wavelengths: 1.59 and 1.683 $\mu$m is shown in Fig.\ref{uranus_limbdark_VLT}, together with the observed reflectivities as a function of $\cos(\theta)$ and the sampled reflectivities at the Gauss-Lobatto quadrature points. In this case the storm cloud was too large and too poorly resolved to be discarded by a median average. Hence, only longitudes to the right of the central meridian in Fig.\ref{uranus_limbdark_VLT} were considered. Finally, the IRTF/SpeX observations were long-slit spectroscopy measurements made with the slit along the central meridian. The observations at 30--40$^\circ$N were made at a zenith angle of $\sim 23^\circ$ which is close enough to one of the five Gauss-Lobatto quadrature points ($23.142^\circ$) to be assigned directly to it. At each of the five zenith angles, the combined measurement vector to be fitted by NEMESIS was comprised of the seven HST/WFC3 observations sampled at that zenith angle together with the VLT/SINFONI observations, sampled at 25 equally-spaced wavelengths over the range 1.436 -- 1.863 $\mu$m. In addition to these five spectra, a sixth spectrum was added at $\theta = 23.142^\circ$, containing the IRTF/SpeX observation at 30 -- 40$^\circ$N, sampled at 25 equally-spaced wavelengths over the range 0.8 -- 1.8 $\mu$m. We then fitted these spectra with a range of models to explore the range of solutions that might be compatible with the observations. The error values on the SpeX and SINFONI spectra were assigned as described by \cite{tice13} and \cite{irwin16}, respectively, and we also included the forward-modelling spectrum of \cite{tice13} to account for insufficiencies in our absorption data, vertical layering, model parameterisation, etc. For the HST/WFC3 data, photometric errors were estimated in the reduction process and are listed in Table \ref{tbl-1}.

\begin{figure}
\epsscale{1.0}
\plotone{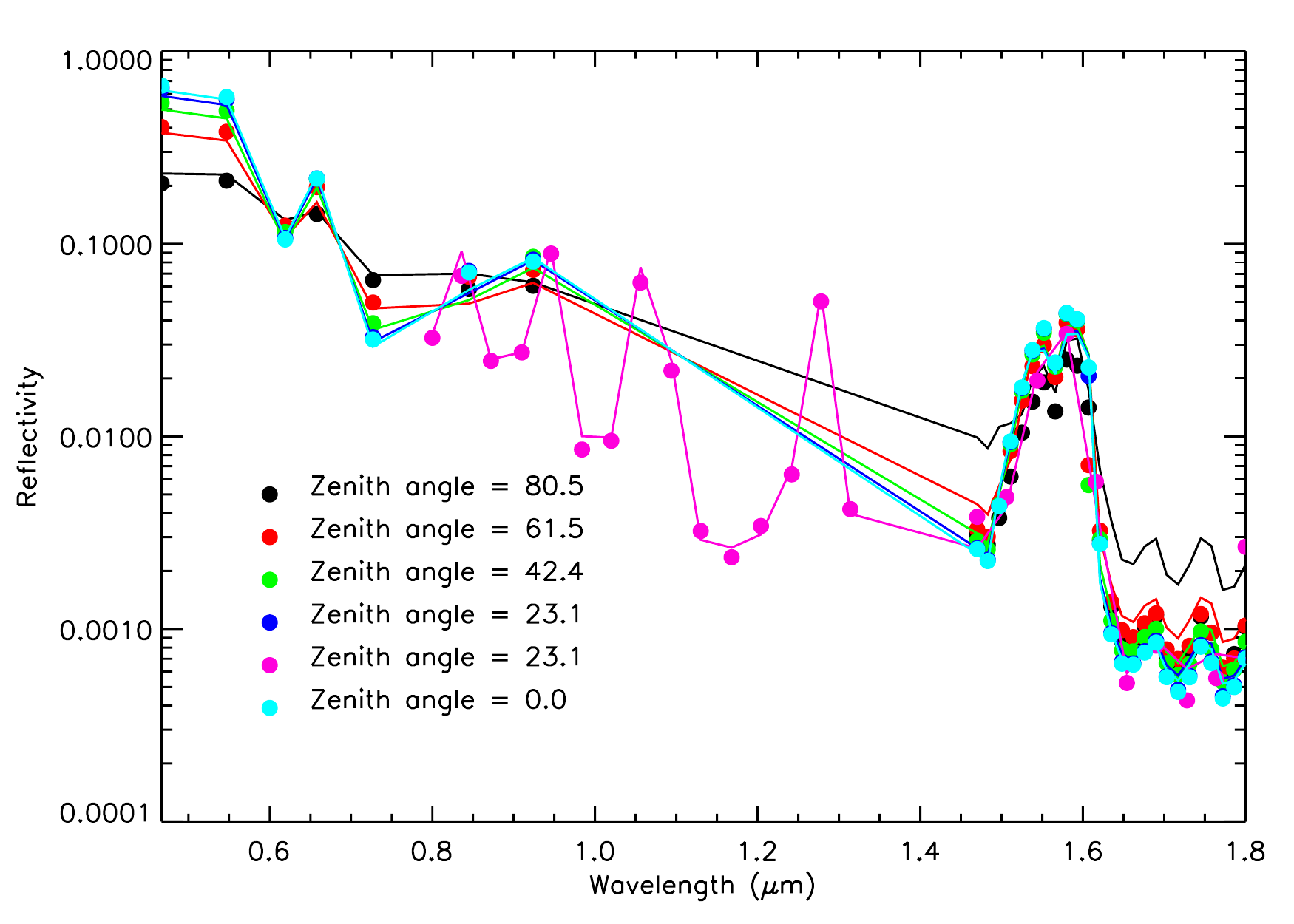}
\caption{Combined WFC3/SINFONI/SpeX limb-darkening observations in the 30 -- 40$^\circ$N latitude range together with the best fit to them with NEMESIS (achieved with an \textit{a priori} imaginary refractive index of 0.01 for the tropospheric cloud and 0.1 for the tropospheric haze, both at all wavelengths). Observations at different zenith angles are shown in different colours and overplotted.   \label{uranus_limbfit1}}
\end{figure}

\begin{figure}
\epsscale{0.7}
\plotone{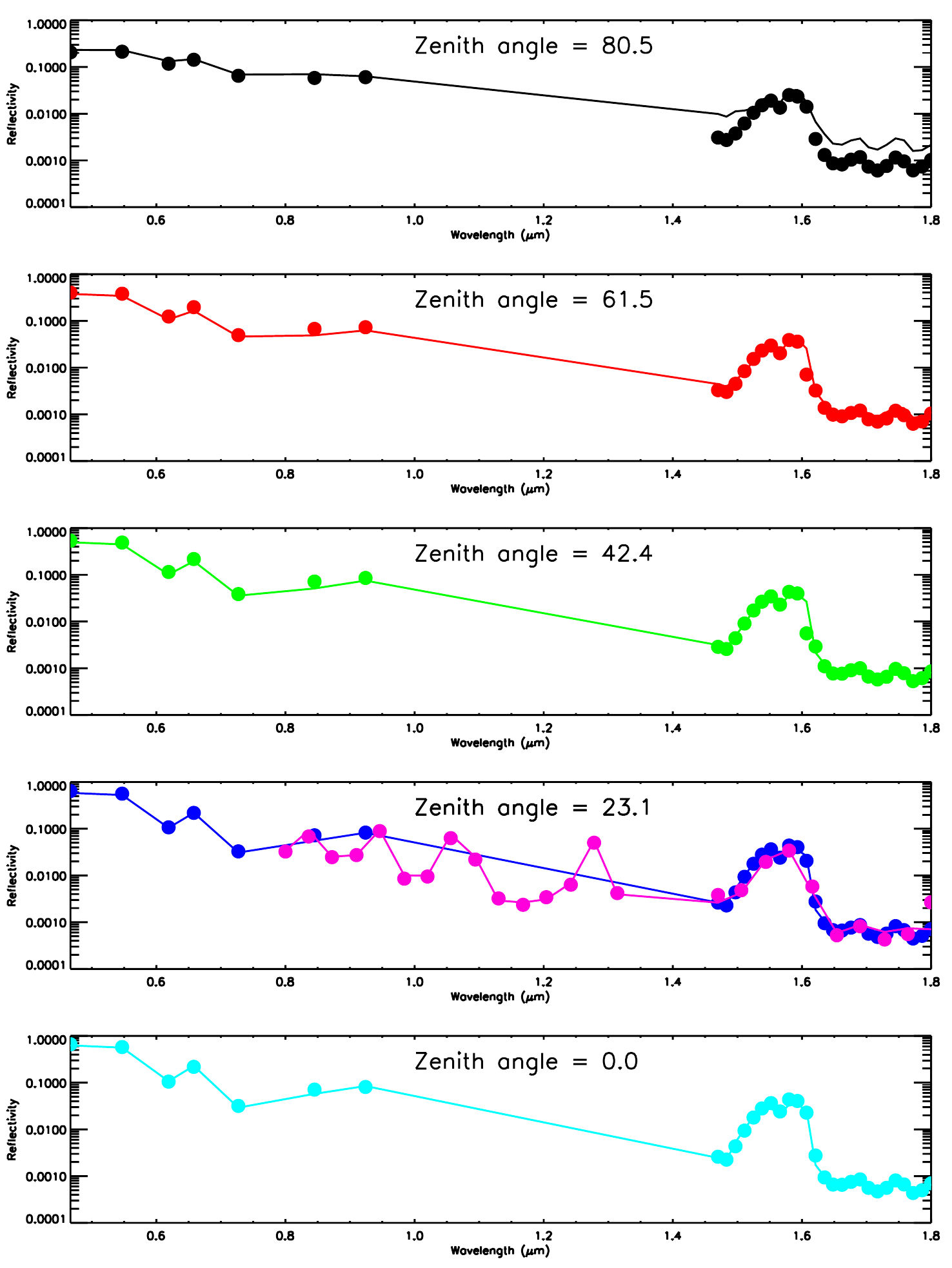}
\caption{Combined WFC3/SINFONI/SpeX limb-darkening observations in the 30 -- 40$^\circ$N latitude range together with the best fit to them with NEMESIS (achieved with an \textit{a priori} imaginary refractive index of 0.01 for the tropospheric cloud and 0.1 for the tropospheric haze, both at all wavelengths). Observations at different zenith angles are shown in different colours and separate panels. For the observations at $23.1^\circ$ the WFC3/SINFONI observations are shown in blue and the SpeX observations are shown in pink.  At high zenith angle ($80.5^\circ$), the broad PSF of VLT/SINFONI includes a contribution from deep space and reduces the reflectivity of the observed data (black points) compared to the model (black curve). \label{uranus_limbfit2}}
\end{figure}

\begin{figure}
\epsscale{1.0}
\plotone{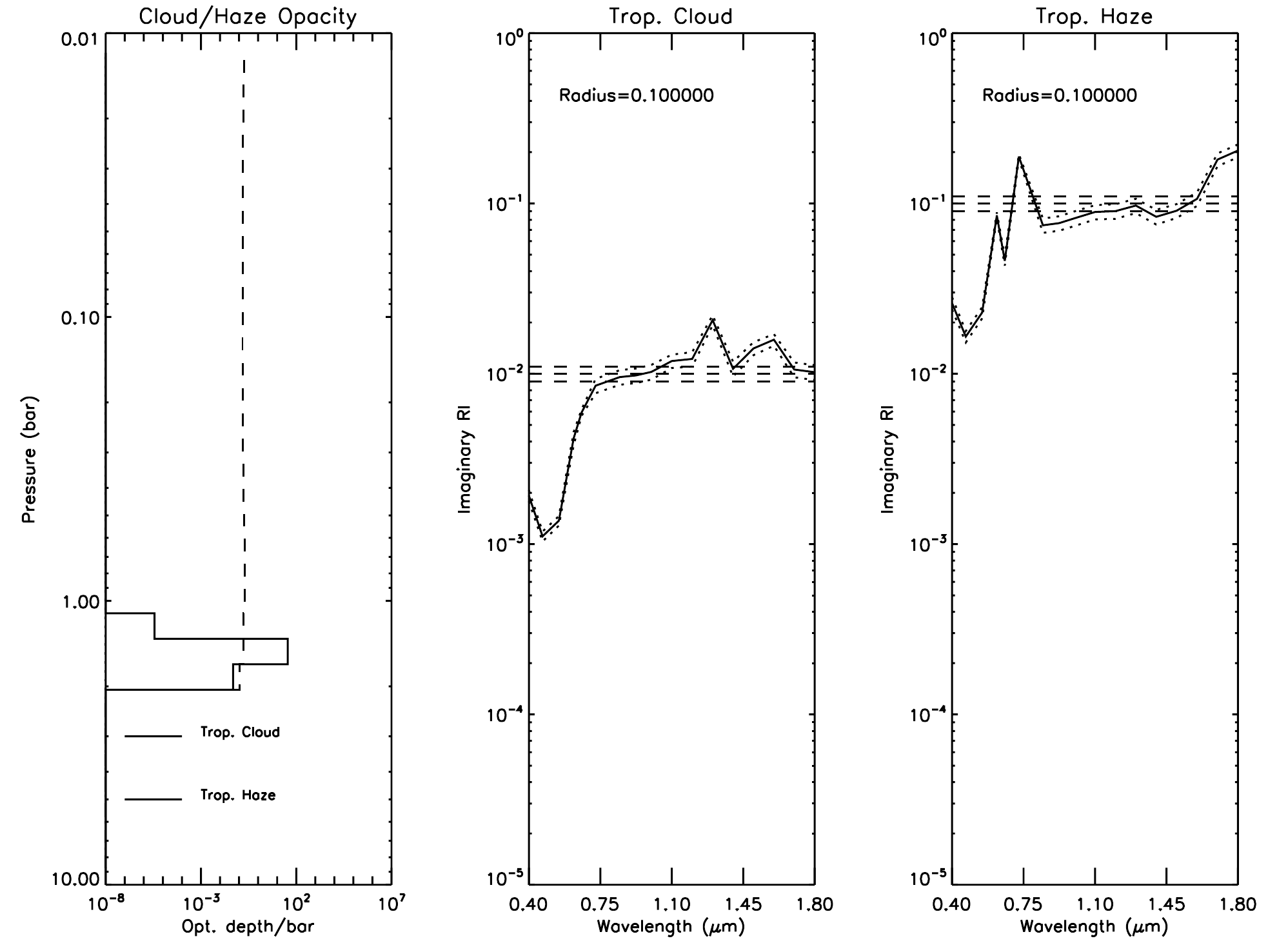}
\caption{Fitted cloud opacity profiles and imaginary refractive index spectra from the combined limb-darkening observations using NEMESIS. Left hand panel shows the retrieved cloud opacity profiles. Remaining panels show the imaginary refractive index spectra retrieved for the tropospheric cloud and tropospheric haze. In these retrievals the methane cloud opacity was fixed at zero and the mean radii of the size distributions was fixed to the values indicated. The horizontal dashed lines in the two right hand panels indicates the \textit{a priori} values (and error) assumed for the imaginary refractive indices.  \label{uranus_limbfit_ret}}
\end{figure}

Once our model was set up we proceeded to test it for varying initial assumptions and found our model to be highly degenerate, with many combinations of initial \textit{a priori} assumptions capable of leading to equally good solutions. In particular we found that there was significant degeneracy between the spectral contribution of the tropospheric haze and the methane cloud, since both give significant reflectivity from the $\sim 1$ bar level, which was difficult to distinguish from each other. We thus assumed that for the background atmospheric state, the methane opacity was low and searched for solutions with just a deep cloud at $\sim$2 bar and a tropospheric haze, varying the assumed particle radii in the two aerosol components and assuming different \textit{a priori} values of the imaginary refractive indices. We found that allowing both the radius of the particles \textbf{and} the imaginary refractive index spectrum to vary allowed too much degeneracy and the model occasionally became unstable (e.g. if two parameters in a retrieval model have similar effect on the modelled spectrum then one can increase and the other decrease to unacceptably large/small values without actually improving the spectral fit). Also, for small particles and wavelengths where the extinction cross-section is in the Rayleigh-scattering (i.e. varying as $1/\lambda^4$), small changes in the particle radius have almost no effect on the modelled spectrum and can fluctuate wildly. Hence, we fixed the mean size of the particles in the tropospheric cloud to be either 1.0 or 0.1 $\mu$m and fixed the mean particle size of the particles in the tropospheric haze to be 0.1 $\mu$m. A series of retrieval tests was run setting the \textit{a priori} imaginary refractive indices of both the cloud and the haze to be 0.001, 0.01 and 0.1, at all wavelengths, giving nine cases in all. Our retrieval setup is summarised on Table \ref{tbl-3}. We also tested different assumptions for the vertical extent of the tropospheric cloud, through varying its fractional scale height, eventually settling at a value of 0.01, which produced a reasonably thin cloud consistent with tests performed with an \textit{a priori} continuous cloud distribution. Our best results were for the case where the tropospheric cloud particles were small (0.1 $\mu$m), and where the \textit{a priori} cloud imaginary refractive indices were 0.01 and the \textit{a priori} cloud imaginary refractive indices were 0.1 (i.e. quite dark); the fitted spectra in this case are shown in Figs.\ref{uranus_limbfit1} and \ref{uranus_limbfit2}, with the retrieved cloud profiles and imaginary refractive index spectra shown in Fig. \ref{uranus_limbfit_ret}. As can be seen, the fits are remarkably good at all wavelengths and viewing geometries. The one exception to this is that the model does not fit the VLT/SINFONI observations well at the highest zenith angle of 80.5$^\circ$. However, the radiances fitted here take no account of the imperfect `seeing' of the VLT/SINFONI observations, which for pixels near the limb (as would be the case for a zenith angle of 80.5$^\circ$, Fig.\ref{uranus_limbdark_VLT}) would lead to mixing of the radiances with cold space, artificially lowering the measured radiances. This problem is less evident for the HST/WFC3 observations because of this instrument's much better spatial resolution. As a result, we increased the errors on the VLT/SINFONI observation at 80.5$^\circ$ to be 100\%, assuming equal mixing between the actual radiance at 80.5$^\circ$ and space, to prevent the model from trying to fit to the spectrum at this angle.

\begin{figure}
\epsscale{1.0}
\plotone{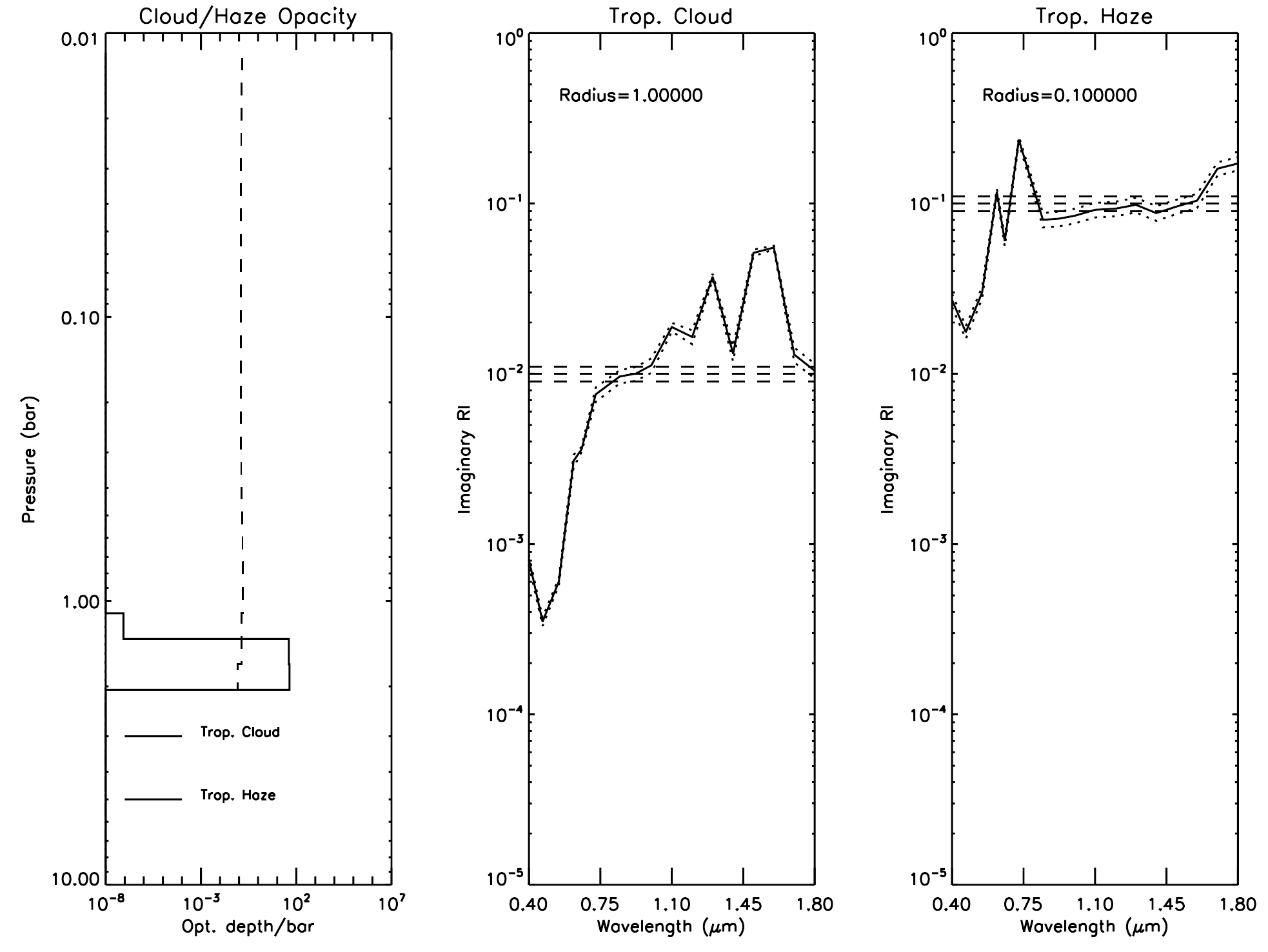}
\caption{As Fig.\ref{uranus_limbfit_ret}, but with the assumed mean radius of the tropospheric cloud particle size distribution increased to 1.0 $\mu$m.   \label{uranus_limbfit_ret_large}}
\end{figure}

\begin{figure}
\epsscale{1.0}
\plotone{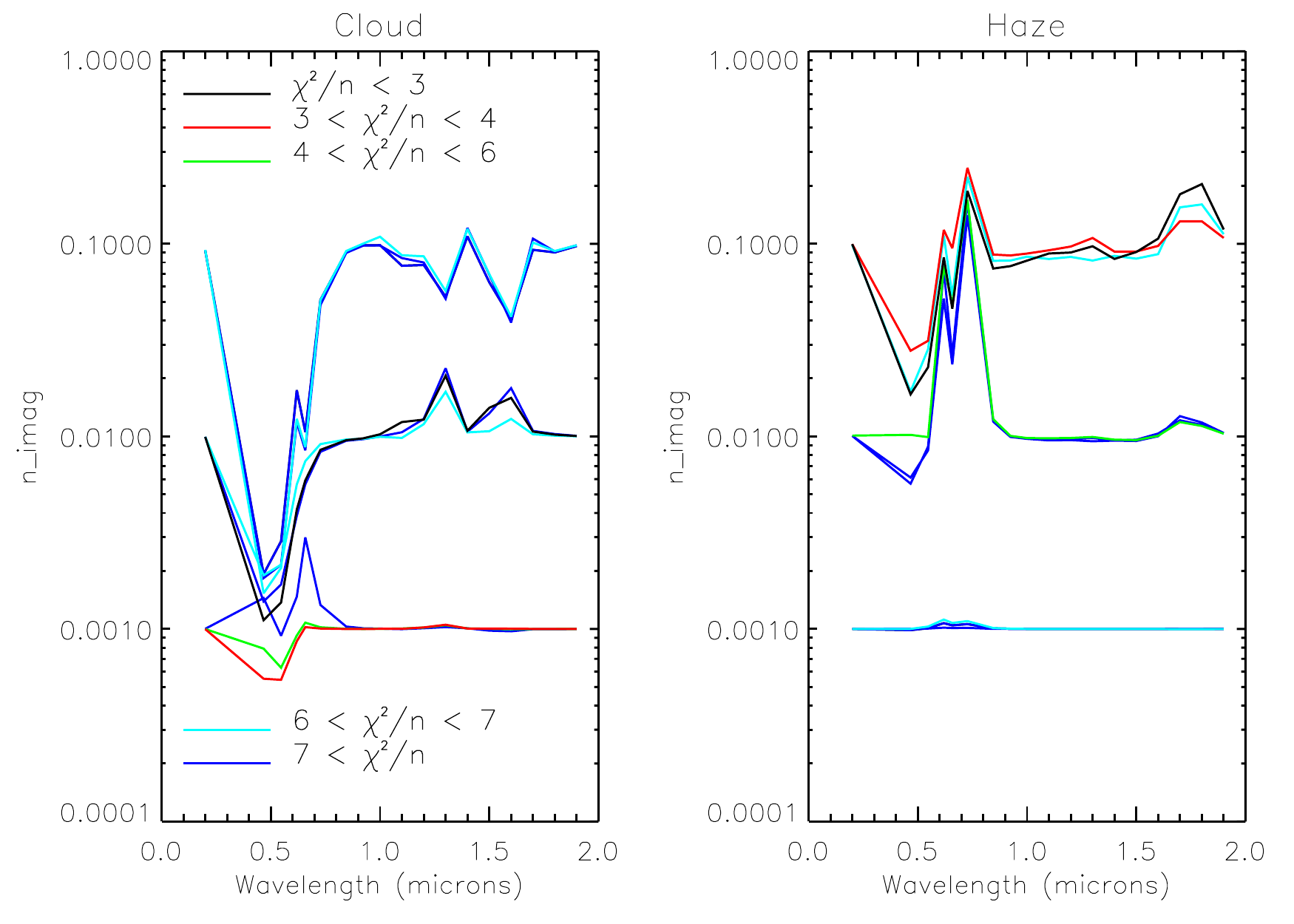}
\caption{Retrieved imaginary refractive index spectra for the tropospheric cloud and haze for the nine different combinations of \textit{a priori} cloud and haze $n_i(\lambda)$. The spectra are colour coded depending on the closeness of fit of the retrieval as indicated.    \label{uranus_compare_nimag}}
\end{figure}

We can see in Fig.\ref{uranus_limbfit_ret} that the imaginary refractive index spectrum of the tropospheric cloud is found to be approximately constant with a value of 0.01 for wavelengths greater than 0.8 $\mu$m, but decreases greatly at shorter wavelengths, becoming less than 0.001 at the shortest observed wavelength of 0.467 $\mu$m, making the cloud particles much more absorbing at longer wavelengths and hence noticeably blue. For the haze, we find that the imaginary refractive index increases longward of 1.6 $\mu$m, has increased absorption  from 0.6 -- 0.8 $\mu$m and is more reflecting at  shorter wavelengths, again making the particles blue in colour. Tests were also done with the \textit{a priori} radius of the deep clouds increased to 1.0 $\mu$m and the best fit retrieval for the case with the same \textit{a priori} imaginary refractive indices is shown in Fig.\ref{uranus_limbfit_ret_large}. Here we see that the haze refractive index spectrum is similar, but that the imaginary refractive index spectrum of the cloud particles varies more significantly with wavelength, reaching $2 \times 10^{-4}$ at the shortest wavelength and increasing with wavelength more noticeably at wavelengths greater than 1.0 $\mu$m. 
The fact that with a relatively simple model such as this we can fit the spectra with either small and large tropospheric cloud particles tells us that we cannot easily distinguish between these possibilities. In both cases the cloud opacity is large and thus we cannot tell if the cloud is indeed a thin discrete cloud, or whether we are simply seeing the top of a much more extended layer below. In terms of scattering it is well known that a cloud of particles with any phase function will have a reflectivity that approaches Lambertian if the opacity is sufficienctly high \citep{plass73} and it would appear that we may be in this situation here. At the longer wavelengths, a tropospheric cloud composed of large particles needs the imaginary refractive indices of the particles to increase with wavelength in order to make their reflectivity drop sufficiently quickly to match the brightness of Uranus' observed reflectance peaks. Conversely, the backscatter of small particles drops naturally with wavelength as $1/\lambda^4$, matching the decreasing reflectance of the peaks quite naturally, without the need to also increase the imaginary refractive index. Since this is a simpler solution we invoke Occam's Razor and use small tropospheric cloud particles in the rest of this paper. Finally, to assess how dependent the inferred imaginary refractive indices might be to the \textit{a priori} starting values,  Fig.\ref{uranus_compare_nimag} compares the retrieved imaginary refractive index spectra of the cloud and haze particles for each of the nine combinations of \textit{a priori} cases, colour coded depending on their goodness of fit. It can be seen that the same trends reveal themselves - the cloud particles need to have lower $n_i$ and thus be more reflective at shorter wavelengths, and the haze particles need to be quite dark to achieve a good fit to the observations.

\subsection{HST Cloud Retrievals}

To test our retrieval model with the HST data alone in the vicinity of the storm cloud features, we first chose six representative locations, noted in Fig.\ref{uranus_false_color2}: 1) centre of the deep cloud; 2) centre of brightest cloud to northeast; 3) centre of second brightest cloud to northeast; 4) a `reference' pixel at the same latitude as cloud 1, but to its east; 5) the small cloud to the southeast of the main cloud; and 6) the trailing bright cloud $100^\circ$ to the east. The radiance spectra at these locations were extracted from the cylindrically mapped observations and fitted with our cloud model. The error estimate of these observations in each filter was again as listed in Table \ref{tbl-1}.
 
\begin{figure}
\epsscale{1.0}
\plotone{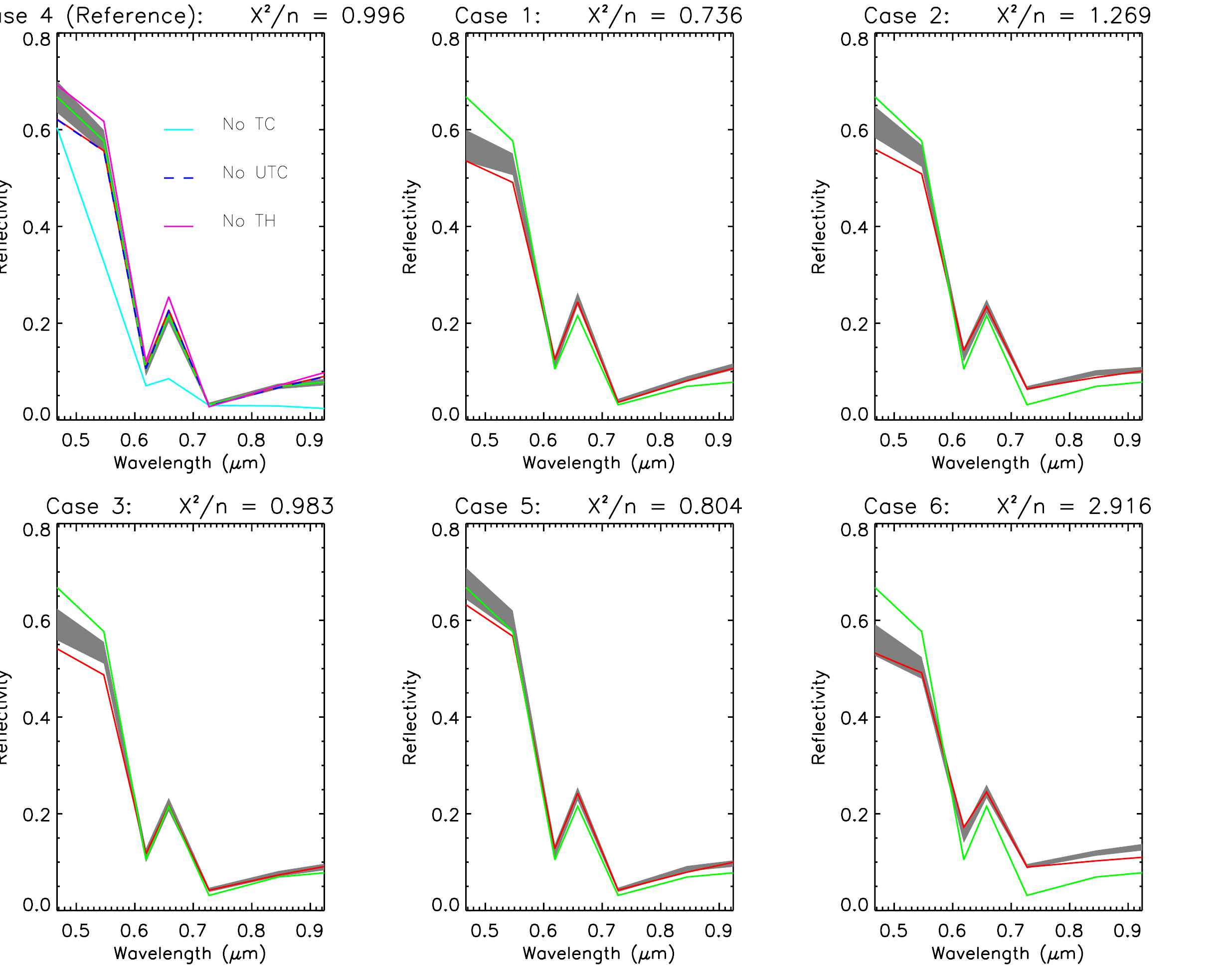}
\caption{Fit to the observed reflectances observed in the six test cases using NEMESIS. In each case the grey region is the measured spectrum and uncertainties, while the red line is the fit. The green lines show the measured spectrum of reference case 4, for ease of comparison.   For the reference case (case 4) shown at top left, the spectra modelled when the opacity of the tropospheric cloud, methane cloud, and tropospheric haze is set to zero is also shown to isolate how the reflectance from each layer contributes to the final modelled spectrum. It should be noted that in this case the contribution from the methane cloud is negligible.\label{case_spectra}}
\end{figure}

\begin{figure}
\epsscale{1.0}
\plotone{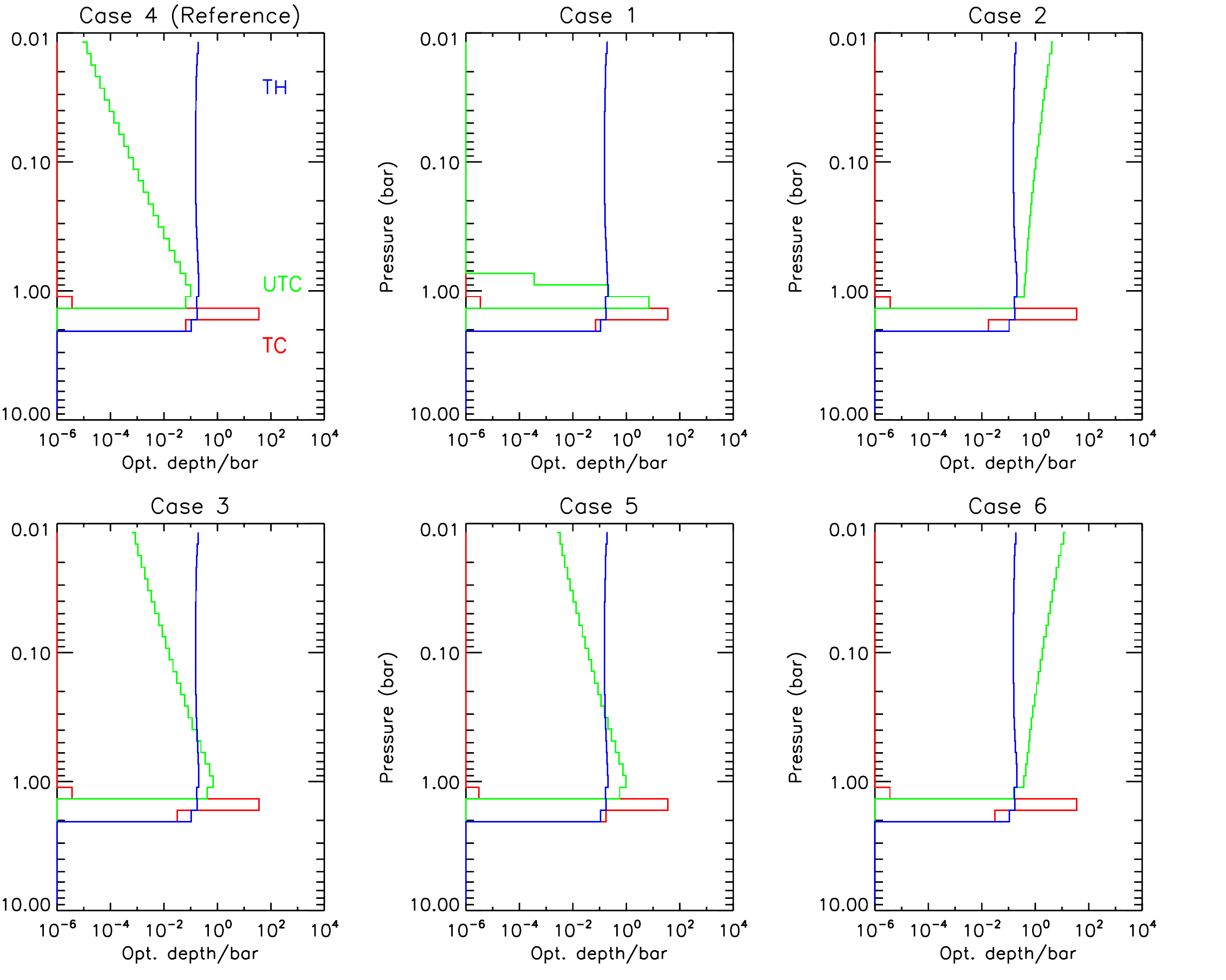}
\caption{Retrieved profiles of cloud opacity from the six test cases using NEMESIS. The tropospheric haze profile and tropospheric cloud profile are fixed to those retrieved from the limb-darkening analysis. In each plot, red indicates tropospheric cloud, blue is tropospheric haze while green is methane cloud opacity.  \label{case_retrievals}}
\end{figure}

Since we could not hope to constrain the vertical level of the tropospheric cloud with these observations, the pressure level was fixed to that retrieved in the limb-darkening study (tests conducted with a variable deep cloud base level proved to be unstable and led to worse fits) of 1.6 bar. We also assumed the tropospheric cloud (TC) opacity and tropospheric haze (TH) opacity were fixed to the best-fitted limb-darkening values of 11.2 and 0.338 respectively (at 467 nm) and thus attempted to fit the HST test case spectra by just retrieving the opacity and fractional scale height of the methane cloud, assumed to be composed of particles of effective radius 0.1 $\mu$m. Tests were also made using methane particles of effective radius 1.0 $\mu$m, but very little difference to the fitting quality to the HST data alone was seen. The size of the TC and TH particles, together with their complex refractive index spectra were similarly held fixed to those values determined by the limb-scattering analysis. Our retrieval model setup is summarised in Table \ref{tbl-4}. Fig. \ref{case_spectra} show the spectra measured in the six test locations, together with our best fit to them with this simple model, while Figs.\ref{case_retrievals} and \ref{case_retrievalsA} show the fitted cloud profiles. Fig. \ref{case_spectra} also shows the modelled spectrum for the reference pixel (case 4) when the opacity of the tropospheric cloud, tropospheric haze or methane cloud are set to zero to show the contribution to the total reflection from these constituents. As can be seen, this simple two-cloud-component model (previously constrained to be consistent with all available limb-darkening observations) can, with the addition of a single methane cloud, parameterised with just two free variables (the total optical depth and vertical extension) , match the observed spectra very well. Hence for Location 1, the main deep cloud, NEMESIS adds an optically thick, but vertically thin cloud, while for Location 2, the bright high cloud to the northeast, the peak opacity is much less, but the cloud more vertically extended, giving a higher abundance of scattering particles in the upper troposphere. 

Given the success of this simple model, we then applied it to the whole area around the clouds, first of all in the area around the main cloud group (Fig. \ref{mainA}) and then also about the trailing cloud to the east (Fig. \ref{mainB}). As can be seen the model matches the observed radiance images very well and returns plausible variations in the thickness and vertical extent of the methane cloud layer, which we have assumed here to be responsible for all the small-scale variation observed. We also find an increased abundance of methane cloud opacity at the northern edge of the region shown in Fig.\ref{mainA}, which accounts for the polar `hood'. 

\subsection{VLT/SINFONI Retrievals}

\begin{figure}
\epsscale{1.0}
\plotone{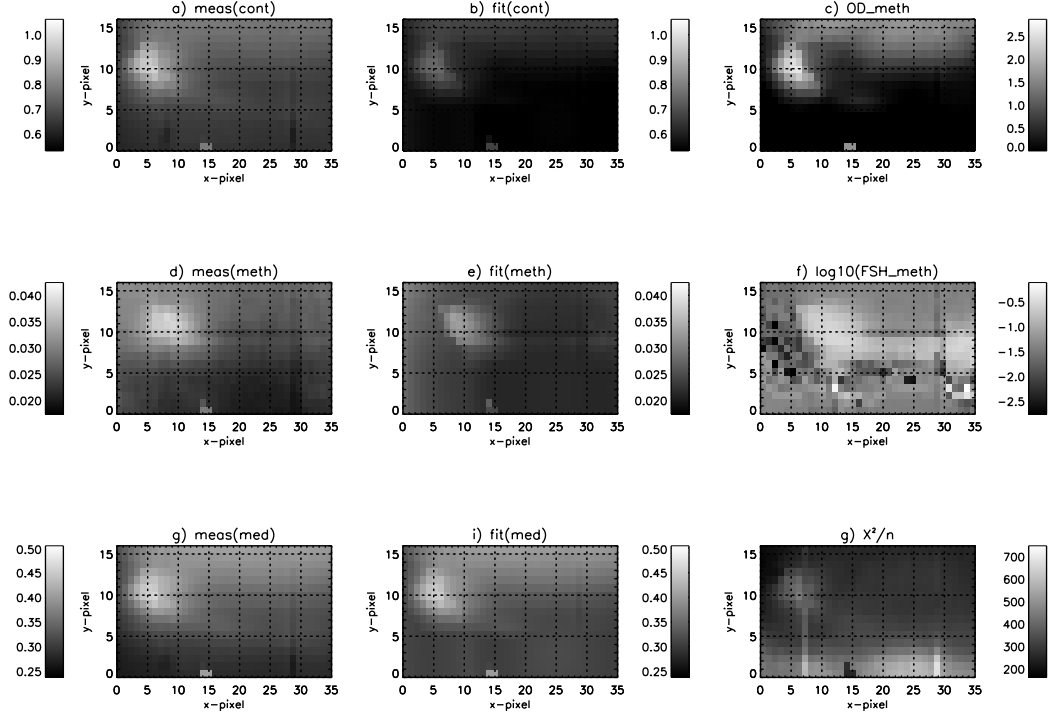}
\caption{Fitted radiances and derived methane cloud profile parameters from region about the main cloud features observed by VLT/SINFONI on 31st October 2014. The figure compares the measured and observed radiances averaged of the wavelength intervals: 1) `cont' (continuum reflection from 1.55--1.62 $\mu$m);  2) `meth' (methane absorbing wavelengths 1.62-1.65 $\mu$m; and c) `med' (medium methane absorption 1.45 - 1.55 $\mu$m) and shows the variation of the fitted variation in the methane opacity and fractional scale height (the latter as a log value to highlight variations). The variation of $\chi^2/n$ is also shown, where the large errors are explained in the text. \label{mainVLT}}
\end{figure}

We also applied our model to fitting the best resolved VLT/SINFONI observations recorded on 31st October 2014, reported by \cite{irwin16}, again fixing the tropospheric cloud and tropospheric haze opacities to the values determined from the limb-darkening study, and allowing only the optical depth and fractional scale height of an additional methane cloud to vary. Figure \ref{mainVLT} shows our resulting fits. Again we see good correspondence between the observed and fitted features at different wavelengths, and again see that the deeper cloud to the southeast is formed by a thick methane cloud of limited vertical extend, while the higher cloud to the northeast is modelled as a thinner methane opacity cloud, but of significant vertical extent. As for the HST wavelengths, we also find an increased abundance of methane cloud opacity at the northern edge of this region, which again accounts for the polar `hood'. The very large values of the $\chi^2/n$ can be understood from the fact that we have used a model fitted to the general limb-darkening observations, fitting to just 25 wavelengths across the VLT/SINFONI range. For the retrievals shown in Fig.\ref{mainVLT} we fitted to spectra sampled at 241 wavelengths and at this level of sampling there are deficiencies in the methane absorption data of \cite{kark10}, as noted by \cite{irwin15} and \cite{irwin16}, who instead used the methane line data of \cite{campargue12}, which was found to model much better the observed spectra in this range. Since our primary purpose in this paper was to model the HST/WFC3 observations, we were constrained to use the methane absorption data of \cite{kark10}, but we can see that our simple model matches the gross features of both the HST/WFC3 and VLT/SINFONI data. We did, however, explore whether a combined use of both data sets could help constrain the size of methane ice particles. Assuming methane ice particles of effective radius 0.1 $\mu$m we determine a peak opacity in the main, deep storm cloud of $\sim1.4$ for HST (Fig.\ref{mainA}) and 2.5 for VLT (Fig.\ref{mainVLT}), a ratio of 0.56. If our model was truly correct, it should be consistent with all available data (although it is worth noting that 10 days had elapsed between these observations). The fact that we need much more opacity of the small methane particles to match the VLT observations at 1.6 $\mu$m, than the HST observations in the visible suggests that the assumed methane particle radius of 0.1 $\mu$m is too small. Hence, we also compared retrievals conducted where the assumed methane particle radius was 1.0 $\mu$m and obtained peak opacities of 1.3 for HST and 0.5 for VLT, a ratio of 2.6. Hence, this comparison indicates that the effective radius of the methane particles is likely to lie somewhere in the 0.1 -- 1.0 $\mu$m range, i.e. $\sim 0.5$ $\mu$m. 

\section{Discussion and Conclusion}

We use a combination of visible data from HST and infrared data from VLT/SINFONI and IRTF/SPEX to determine the nature and distribution of cloud particles in and around bright features in the 30--40$^\circ$N latitude range (planetographic). Many results of our retrievals are consistent with prior work. We find that a non-convective origin is most consistent with our observations of the cloud features at 30--40$^\circ$N.

\begin{figure}
\epsscale{1.0}
\plotone{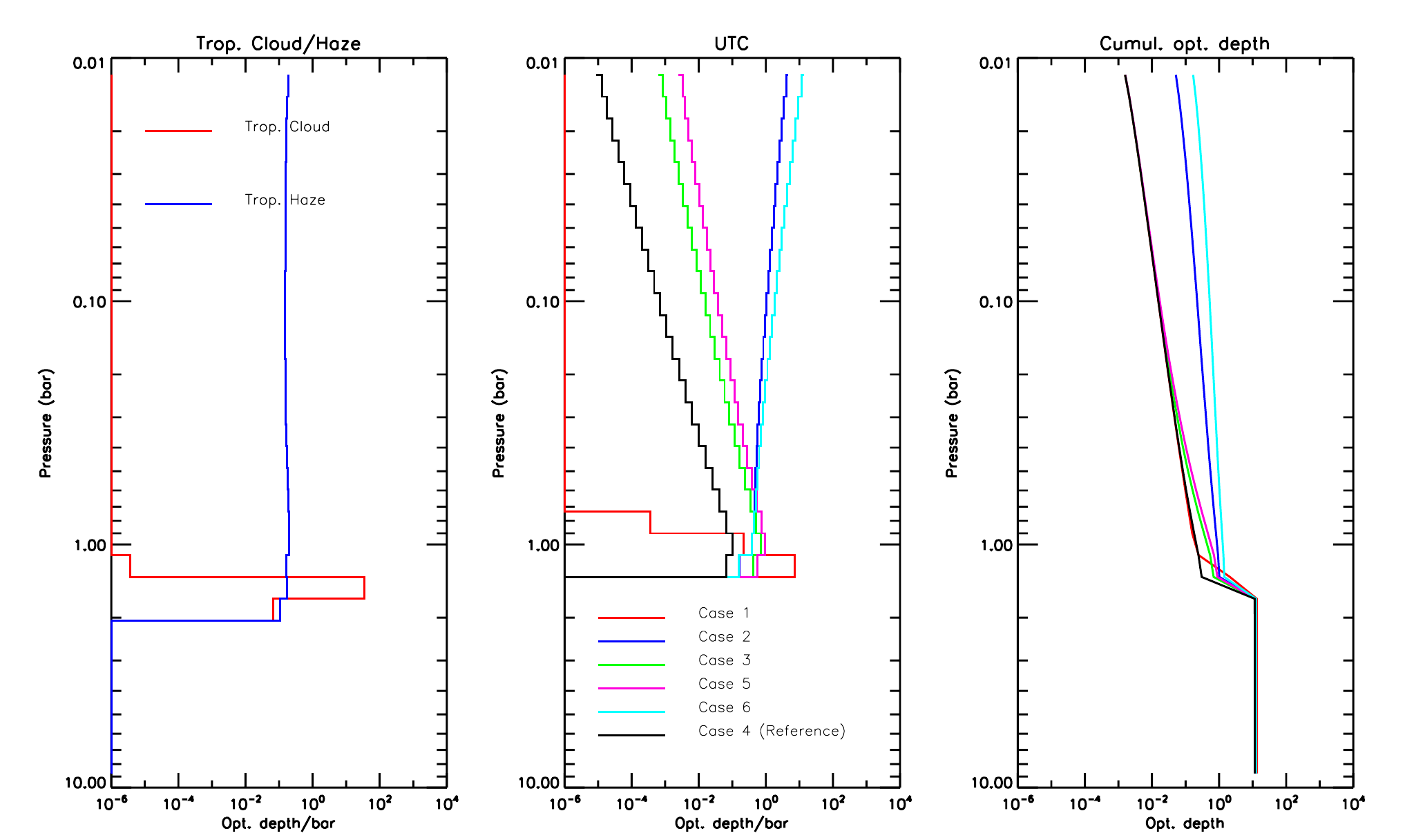}
\caption{Retrieved profiles of cloud opacity (at 457 nm) from the six test cases using NEMESIS, separated by cloud type to show the  profiles of tropospheric cloud and tropospheric haze determined from the limb-darkening analysis and then fixed and the retrieved profile the methane cloud (UTC) for each case. Also shown is the cumulative cloud optical depth retrieved for each case, again at 457 nm. \label{case_retrievalsA}}
\end{figure}

{\em Mean aerosol distribution at 30--40$^\circ$N.} Our simplified model includes a tropospheric haze layer (because its base lies in the troposphere), but the layer extends up to 10 mbar, overlapping with stratospheric aerosol layers taken to be distinct from tropospheric populations in prior works (e.g., \cite{baines95}, \cite{kark09}). Like these prior works, we fit the data with small (0.1 $\mu$m) haze particles. However, the colour of the haze material (as given by the imaginary index of refraction spectrum), is opposite to what was found in \cite{kark09}, who found conservative scattering longward of 600 nm, and increasingly strong absorption going to shorter wavelengths. Our retrievals, for both cloud and haze particles, find a minimum of absorption in the blue (467 nm), and increasing absorption to about 800 nm. At longer wavelengths, $n_i$ is constant for the cloud particles, but increases somewhat for haze particles (Figs. \ref{uranus_limbfit_ret}--\ref{uranus_compare_nimag}). The entire region is well-fit by an optically thick cloud concentrated near 1.6 bar, at the base of the tropospheric haze.  The optical depth per bar (Fig. \ref{case_retrievalsA}) jumps by a factor of about 100 as pressure increases into this cloud layer, similar to the increase in aerosol opacity at the 1.2-bar level in the retrieval of \cite{kark09}. Most other prior analyses focused only on aerosol distributions in discrete cloud features.

For the generally observed spectra throughout the 0.467--1.8 $\mu$m region, we found the analysis to be surprisingly degenerate. Even when simultaneously fitting observations at a range of zenith angles we find the the observations can be fit with a wide range of cloud models. To limit the parameter range, physically-constrained models incorporating microphysics and dynamics could be used. Further observations of such clouds at higher spectral resolution would also be advantageous.

{\em Aerosols to the north and south.} We can also see in Fig. \ref{mainA} (panel C) that enhanced, vertically restricted methane cloud opacity is found a latitudes north of about 38$^\circ$, at the edge of the `polar hood.' More generally, we have found that the background distribution of particles in Uranus' atmosphere at these latitudes is well-matched with a vertically thin, but optically thick cloud of reasonably absorbing particles (except at the shorter wavelengths) in a tropospheric cloud, overlain by a vertically extended tropospheric haze of infrared-absorbing particles.

South of about 30$^\circ$N, in both our HST area retrievals (Fig. \ref{mainA}) and VLT area retrievals (Fig. \ref{mainVLT}), values of reduced $\chi^2$ are much higher than in other areas. This clearly indicates that there is a significant difference in the background atmosphere to the north and south of a boundary at ~30$^\circ$N. Quantifying this difference would require detailed modelling beyond the scope of our investigation of discrete cloud features. The work of \cite{kark09} suggests that CH$_4$ concentration or aerosols at $p > 2$ bar could be responsible for the difference. In either case, the 30$^\circ$N latitude marks some kind of dynamical boundary, but cloud tracers to date have not been able to conclusively demonstrate a corresponding feature in the zonal wind field \citep{hammel09, sromovsky12b}.

{\em Ruling out convection.} Two factors initially might suggest a convective origin for the cloud features: their high reflectivity, and the suggestion of sheared structure similar to thunderstorm anvils on Earth, or the 2010 Saturn storm. A closer look at these factors shows they are not consistent with a convective origin.

Cloud features with the strongest contrast with the surroundings---such as features 2 and 6 in Fig. \ref{uranus_false_color2}, or features at other latitudes (\cite{dePater15}; or \cite{sromovsky07})---are bright relative to their surroundings mainly because of the relatively high atmospheric absorption at long wavelengths. Our modelling of features 2 and 6 (Fig. \ref{case_retrievalsA}) explains them as concentrations of high-altitude particles, but they are still much less optically thick than the tropospheric cloud deck near 1.6 bar. Vigorous moist convection (at least on other planets) produces clouds that are optically thick at multiple wavelengths sampling the full vertical extent of the cloud \citep{gierasch00}. Thus, although these features are bright relative to their surroundings, their actual optical depths ($\sim 1$ at 0.467 nm) are not as large as would be expected for convective features. For a convective plume, optical depths would be essentially infinite at all wavelengths.

\begin{figure}
\epsscale{0.9}
\plotone{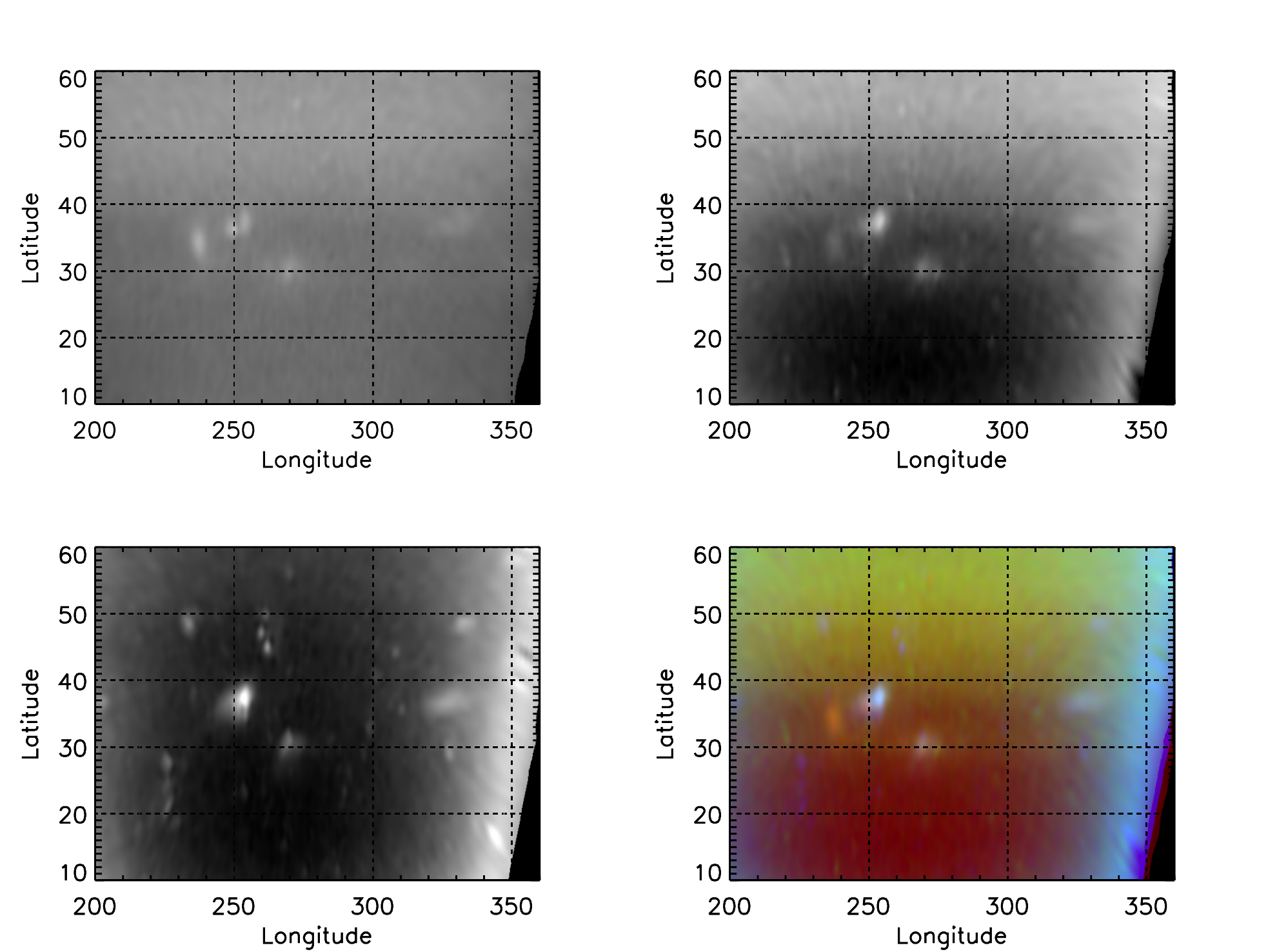}
\caption{Observations on 9th November 2014 near 01:00 UT in the region of the storm feature in: 1) F924N (weak methane absorption - top left); 2) FQ619N (medium methane absorption - top right);  3) FQ727N (strong methane absorption - bottom left) filters; and 4) colour composite (bottom right) using the same colour scheme as Fig.\ref{uranus_false_color}. There are some cosmic ray defects in the FQ727N image, but we see that the main storm cloud (240$^\circ$E, 34$^\circ$N) is only clearly seen in the F924N filter (and thus appears red in the colour composite), indicating that it lies in, or only just above the main cloud deck at $\sim$ 2 bars. This feature is accompanied by a high, bright cloud at 250$^\circ$E, 37$^\circ$N and another second high cloud at 270$^\circ$E, 30$^\circ$N. Both these latter clouds appear white in the false colour composite, showing them to be high. Finally, there is a general increase in opacity of the upper clouds polewards of 40$^\circ$N, with the false colour image generally  changing from red to green as we cross the 40$^\circ$N latitude, indicating the latitudinal boundary of the polar `hood'. \label{uranus_false_color1}}
\end{figure}

\begin{figure}
\epsscale{1.0}
\plotone{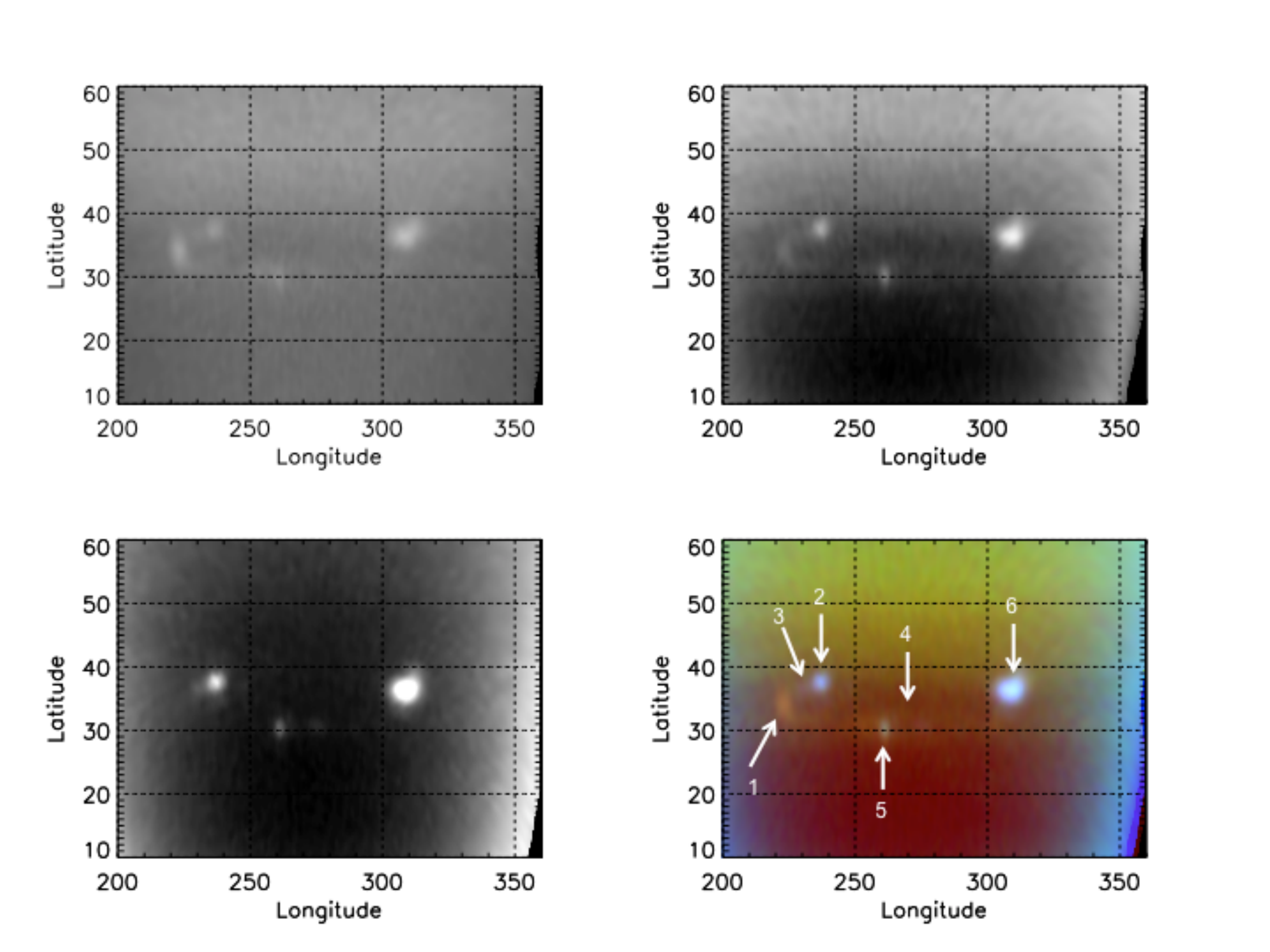}
\caption{Observations on 9th November 2014 near 18:30 UT in the region of the storm feature in F924N (deep), FQ619N (medium) and FQ727N (high) filters, and colour composite (as Fig.\ref{uranus_false_color1}). In this case, the images are rather clearer and less affected by defects. We can see that the high cloud at 37$^\circ$N to the northeast of the main deep feature is actually two clouds, with the upper right part being brighter. It is also apparent in these observations that the three clouds are linked by a curved line of opacity that stretches from the upper right cloud, through the main deep cloud and then trails away to the right, approaching an asymptote along the 30$^\circ$N latitude line. Far to the east of the main feature, at 310$^\circ$E, 36$^\circ$N is a second very high cloud, which as it moves together with the main storm cloud (Fig.\ref{uranus845summary}) might perhaps be part of the same atmospheric system. Also indicated in this figure are the locations of the six test regions chosen to constrain the setup of our retrieval model.     \label{uranus_false_color2}}
\end{figure}

\begin{figure}
\epsscale{1.0}
\plotone{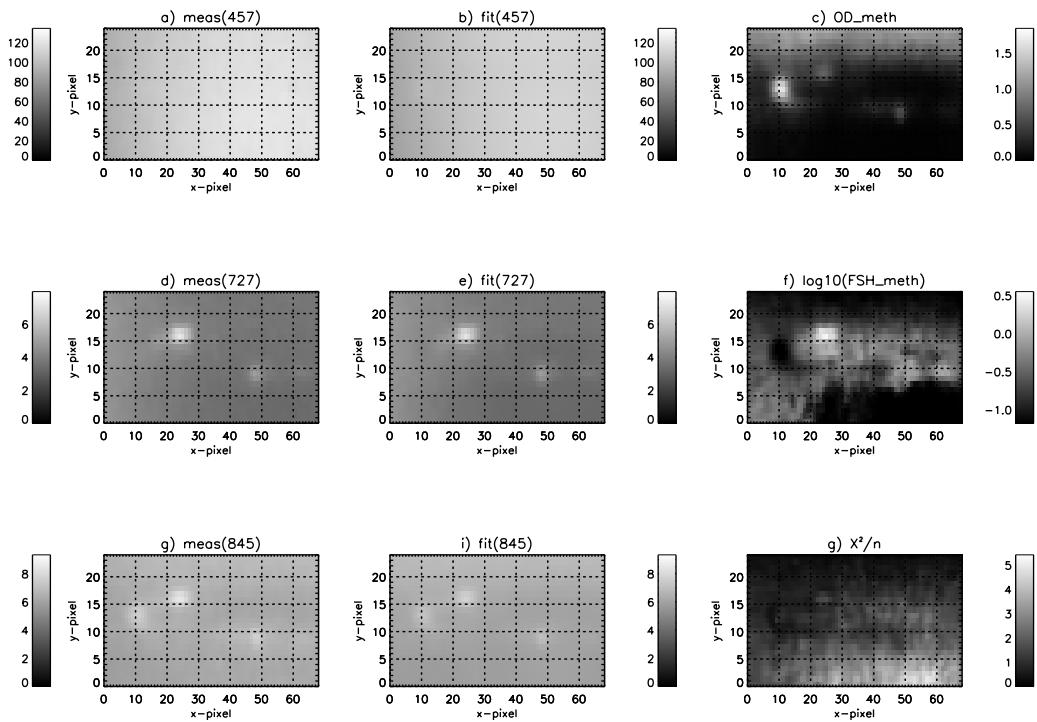}
\caption{Fitted radiances and derived methane cloud profile parameters from region about the main cloud features (encompassing features 1 -- 5 from Fig. \ref{uranus_false_color2}). The figure compares the measured and observed radiances at 0.457, 0.727 and 0.845 $\mu$m, respectively and shows the variation of the fitted variation in the methane opacity and fractional scale height (the latter as a log value to highlight variations). The variation of $\chi^2/n$ is also shown. \label{mainA}}
\end{figure}

\begin{figure}
\epsscale{1.0}
\plotone{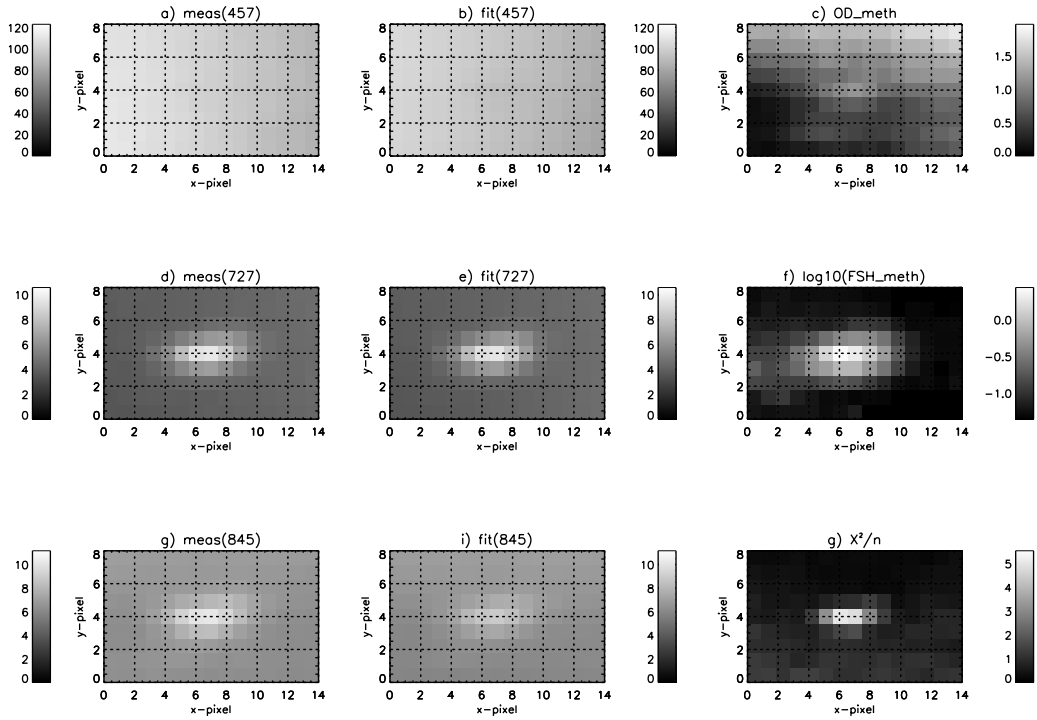}
\caption{As Fig. \ref{mainA}, but centred about the trailing high cloud to the east (i.e. feature 6 of Fig. \ref{uranus_false_color2}).  \label{mainB}}
\end{figure}

Sheared structure was originally invoked to explain our VLT/SINFONI observations \citep{irwin16}. \cite{dePater15} also drew a parallel between the convective superstorm on Saturn in 2010, and a streak/streamer/tail similar to the ones attached to feature 1 in Figs. \ref{uranus_false_color1} and \ref{uranus_false_color2}. Sheared structure is produced when a highly localised convective source lofts particles to a stopping altitude somewhat higher than the tropopause, where wind shear then advects the particles away from the source. However, both \cite{dePater15} and our analysis find that the streamers seen in 2014 did not fit this profile, due to their relatively low altitude between 1 and 2 bar. 

Cumulonimbus anvils form because storm energies provide a high upward flux of particles, until a stopping point is abruptly reached within a strong gradient of increasing static stability with altitude. But the 1--2 bar level in Uranus' atmosphere is only weakly stratified \citep{lindal87, hammel09}. Under the simplest assumption of a linear vertical wind shear gradient, a convective plume terminating in the 1--2 bar range would loft particles to a range of altitudes spanning several bars, sensing a range of wind shear, resulting in a streamer whose brightness distribution should depend strongly on distance from the source. In comparison, the 2014 streamers maintained a more or less constant reflectivity for several thousand kilometres. Additionally, \cite{dePater15} suggested that southward advection of aerosols, responding to the mean zonal wind profile, would explain the streamer extending eastward from the source. However, this mechanism cannot explain the morphology of the streamers attached to feature 1 in Figs. \ref{uranus_false_color1} and \ref{uranus_false_color2}. In our imaging data, streamers extend eastward both to the south AND north of feature 1. If horizontal advection within the mean zonal wind field is invoked, the northern streamer should instead extend to the west.

The feature 1 aerosol profile (Fig. \ref{case_retrievalsA}) is sharply peaked at the methane cloud base, which is fixed at the 1.23-bar level. But with only seven HST wavelengths it is quite possible that we could match all the discrete feature data just as well by the addition of a vertically thin cloud with variable base pressure in the 1.2--0.1 bar pressure region, or in the case of feature 1, with a greater opacity in the 1.6-bar main tropospheric cloud deck. To differentiate between these possibilities would require spectral resolution sufficient to measure the shape of the methane absorption peaks as well as their depth. Our VLT/SINFONI observations may have sufficient resolution to be able to differentiate between these possibilities and we are currently reanalysing these data. However, this work is too premature to report here.

Convective plumes have been intensively studied on Jupiter and Saturn, but never conclusively identified on Uranus or Neptune. In Jupiter and Saturn, convective storms are probably driven (or at least accelerated) by water and its latent heat. The predominance of water as the working fluid in moist convection is consistent with lightning correlated with convective storms, the higher solar abundance of water compared to other volatiles, and the fact that water condensation happens at the deepest levels, where densities (and thus potential energy) are highest. But the water cloud level in Uranus and Neptune \citep{atreyawong05} is some 3--5 scale heights below the observable atmosphere. Most likely, convective storms would not be directly observable, although their energetic effects could lead to observable secondary
convection in CH$_4$ or H$_2$S cloud layers.

{\em The vortex companion scenario.} With convection being broadly inconsistent with the observations, it seems plausible that the bright features may be companion clouds orographically produced by a deep anticyclonic vortex. A dark spot was indeed seen in 2006 near 28$^\circ$N \citep{hammel09}, accompanied by bright cloud features to the north. Bright cloud features have also been seen in 1998--1999 \citep{sromovsky00}, 2004--2005 \citep{sromovsky07}, and 2007 \citep{sromovsky09}. The features could be related to a long-lived vortex, or to multiple vortices, with formation conditions somehow being highly favorable near 28$^\circ$N.

The distribution of cloud features (Figs. \ref{uranus_false_color1} and \ref{uranus_false_color2}) in an approximate east-west-extended ellipse actually draws a parallel to a different type of vortex: cyclones on Jupiter. Cyclones have the opposite sense of rotation to anticyclones such as the Great Dark Spot on Neptune \citep{smith89} or the Great Red Spot on Jupiter. Jovian cyclones are often marked by turbulent, active convection around their periphery, and can often have east-west/north-south diameter ratios similar to the shape of the ellipse outlined by features 1--6. At this point, the resemblance between the observed features and jovian cyclones is qualitative, and substantial new observations would be needed to test such a scenario. These observations could include local wind measurements conducted by an orbiter.

{\em Future work.}In the future, it is apparent that ground-based (or observation from space telescopes in orbit about the earth) observations of Uranus' clouds suffer from a significant degeneracy of their solutions. We lack, at the moment, sufficient laboratory measurements to physically characterise the scattering properties of the likely cloud constituents, which are likely in any case to not be pure condensates at all, but coated or mixed with photochemical products settling down from above, similar to processes in Jupiter's atmosphere that may be responsible for the scarcity of ammonia ice spectral signatures (e.g., \cite{atreya05}, \cite{kalogerikas08}.

We also lack enough temperature observations to sufficiently constrain general circulation dynamical models (GCMs) to establish the overall circulation of the atmosphere and model how it varies with seasons, giving rise to instabilities at mid-latitudes in the periods around the equinoxes that allow discrete clouds, such as those analysed here, to form. It may be that the best way to clarify these and many other uncertainties in our understanding of the Uranus' atmosphere will be to have a dedicated space mission in the future, including, if possible, entry probes to measure \textit{in situ} the properties of Uranus' atmosphere and clouds and orbiters to better constrain the detailed flow field in the atmosphere. Such missions, following on from the spectacularly successful Galileo mission to Jupiter and Cassini mission to Saturn, would advance enormously our understanding of the atmosphere of Uranus and establish why it is so different to that of any other giant planet. This would be of benefit not only to solar system scientists, but also to exoplanetary scientists as they discover and characterise the atmospheres of ever cooler exoplanets.

\section{Acknowledgements}

The VLT/SINFONI observations were performed at the European Southern Observatory (ESO), Proposal 092.C-0187. This work was based on observations made with the NASA/ESA Hubble Space Telescope under programs GO13937/14334. Support for this program was provided by NASA through a grant from the Space Telescope Science Institute, which is operated by the Association of Universities for Research in Astronomy, Inc., under NASA contract NAS5-26555.  Patrick Irwin and Daniel Toledo acknowledge the support of the UK Science and Technology Facilities Council. The authors wish to recognise and acknowledge the very significant cultural role and reverence that the summit of Mauna Kea (where IRTF is located) has always had within the indigenous Hawaiian community. We are most fortunate to have the opportunity to conduct observations from this mountain.  Glenn Orton was supported by a grant from NASA to the Jet Propulsion Laboratory, California Institute of Technology. 



{\it Facilities:}  \facility{HST (WFC3)}, \facility{VLT (SINFONI)}, \facility{IRTF (SpeX)}.

\clearpage




\clearpage

\clearpage

\clearpage

\clearpage

\clearpage

\clearpage

\clearpage

\clearpage

\clearpage

\clearpage

\clearpage

\clearpage

\clearpage

\clearpage

\clearpage

\clearpage

\clearpage

\clearpage

\clearpage

\begin{table}
\begin{center}
\caption{HST/WFC3 Filters used in 2014 Uranus OPAL observations and estimated photometric errors.\label{tbl-1}}
\begin{tabular}{llll}
\tableline
Name (Aperture) & $\lambda_0$ (nm) & FWHM (nm) & Photometric\\
 & &  & error(\%) \\
\tableline
F467M (UVIS2) & 467 & 21.5 & 5.1 \\
F547M (UVIS2) & 547 & 70.9 & 5.1 \\
FQ619N (UVIS) & 619 & 6.1 & 6.0 \\
F658N (UVIS2) & 658 & 2.8 & 5.1 \\
FQ727N (UVIS) & 727 & 6.5 & 6.5 \\
F845M (UVIS2) & 845 & 87.6 & 5.1 \\
FQ924N (UVIS) & 924 & 8.9 & 6.5 \\
\tableline
\end{tabular}
\end{center}
\end{table}

\begin{table}
\begin{center}
\caption{Five-point Gauss-Lobatto quadrature points and weights.\label{tbl-2}}
\begin{tabular}{lll}
\tableline
Zenith Angle  ($\theta^\circ$) & $\mu = \cos(\theta)$ & Weight \\
\tableline
0.00000 & 1.000000 & 0.022222 \\
23.1420 & 0.919534 &  0.133306 \\
42.3729 & 0.738774 & 0.224889 \\
61.4500 & 0.477925 & 0.292043 \\
80.4866 & 0.165279 & 0.327540 \\
\tableline
\end{tabular}
\end{center}
\end{table}

\begin{table}[]
\centering
\caption{Retrieval set up for limb-darkening analysis of HST, VLT and IRTF observations.}
\label{tbl-3}
\begin{tabular}{|l|l|l|}
\hline
          & Cloud Properties   & Cloud Scattering Properties \\   \cline{1-3}                                                                                                                                                                        
TC      & \begin{tabular}[c]{@{}l@{}}Variable base pressure.\\ Variable opacity.\\ FSH fixed to 0.01.\end{tabular}   & \begin{tabular}[c]{@{}l@{}}Effective radius of size distribution set to \\ 0.1 $\mu$m or 1.0 $\mu$m with variance 0.1.\\ \textit{A priori} $n_i$ set to 0.001, 0.01, 0.1 \\ at all wavelengths and retrieved.\\ $n_r$ set to 1.4 at 467 nm.\end{tabular} \\ \hline
UTC & \begin{tabular}[c]{@{}l@{}}Base set by condensation level \\to be 1.23 bar.\\ Opacity set to zero.\\ FSH fixed to 0.1.\end{tabular} & \begin{tabular}[c]{@{}l@{}}Effective radius of size distribution set to \\ 0.1 $\mu$m or 1.0 $\mu$m with variance 0.1\\ Complex refractive index data \\ of \cite{martonchik94} used.\end{tabular}         \\ \hline
TH            & \begin{tabular}[c]{@{}l@{}}Base pressure locked to TC.\\ Variable Opacity.\\ FSH fixed to 1.0.\end{tabular}                             & \begin{tabular}[c]{@{}l@{}}Radius of size distribution set to \\ 0.1 $\mu$m only with variance 0.1\\ \textit{A priori} $n_i$ set to 0.001, 0.01, 0.1 \\ at all wavelengths and retrieved.\\ $n_r$ set to 1.4 at 467 nm.\end{tabular} \\ \hline          
\end{tabular}
\end{table}

\begin{table}[]
\centering
\caption{Retrieval set up for HST area retrievals.}
\label{tbl-4}
\begin{tabular}{|l|l|l|}
\hline
  & Cloud Properties   & Cloud Scattering Properties \\     \hline                                                                                                                                                                   
TC      & \begin{tabular}[c]{@{}l@{}}Base fixed to 1.6 bar. \\ Opacity (at 467 nm) fixed to 11.2. \\ FSH fixed to 0.01.\end{tabular}          & \begin{tabular}[c]{@{}l@{}}Effective radius of size distribution set to \\ 0.1 $\mu$m or 1.0 $\mu$m with variance 0.1.\\ $n_i$ spectrum set to that retrieved from \\ limb-darkening analysis (\textit{a priori} $n_i$ = 0.01).\\ $n_r$ set to 1.4 at 467 nm.\end{tabular} \\ \hline
UTC & \begin{tabular}[c]{@{}l@{}}Base set by condensation level \\ to be 1.23 bar.\\ Variable Opacity. \\ Variable FSH.\end{tabular} & \begin{tabular}[c]{@{}l@{}}Effective radius of size distribution set to \\ 0.1 $\mu$m or 1.0 $\mu$m with variance 0.1\\ Complex refractive index data \\ of \cite{martonchik94} used.\end{tabular}   \\ \hline
TH            & \begin{tabular}[c]{@{}l@{}}Base locked to that of TC at 1.8 bar.\\ Opacity fixed to 0.338. \\ FSH fixed to 1.0.\end{tabular}                             & \begin{tabular}[c]{@{}l@{}}Effective radius of size distribution set to \\ 0.1 $\mu$m only with variance 0.1.\\ $n_i$ spectrum set to that retrieved from \\ limb-darkening analysis  (\textit {a priori} $n_i$ = 0.1).\\ $n_r$ set to 1.4 at 467 nm.\end{tabular} \\ \hline            
\end{tabular}
\end{table}

\end{document}